\documentclass[longauth]{aa} 
\usepackage{graphicx}
\usepackage{txfonts}
\usepackage{natbib}

\def\eg{{\it e.g.\ }}

\def\ie{{\it i.e.\ }}
\def\vs{{\it versus\ }}

\def\thnn{$^{th}$ n.n. }

\begin{document}
   \title{The zCOSMOS 10k-sample\thanks{European Southern Observatory (ESO), Large Program 175.A-0839}: the role of galaxy stellar mass in the colour-density relation up to $z\sim1$}


\author{
O.~Cucciati\inst{1,2}
\and
A.~Iovino\inst{2}
\and
K.~Kova\v{c}\inst{3}
\and
M.~Scodeggio\inst{4}
\and
S.J.~Lilly\inst{3}
\and
M.~Bolzonella\inst{5}
\and 
S.~Bardelli\inst{5}
\and
D.~Vergani\inst{5}
\and
L.A.M.~Tasca\inst{1,4}
\and
E.~Zucca\inst{5}
\and
G.~Zamorani\inst{5}
\and
L.~Pozzetti\inst{5}
\and
C.~Knobel\inst{3}
\and
P.~Oesch\inst{3}
\and 
F.~Lamareille\inst{6}
\and
K.~Caputi\inst{3}
\and
P.~Kampczyk\inst{3}
\and
L.~Tresse\inst{1}
\and 
C.~Maier\inst{3}
\and  
C.M.~Carollo\inst{3}
\and
T.~Contini\inst{6}
\and
J.-P.~Kneib\inst{1}
\and
O.~Le~F\`{e}vre\inst{1}
\and
V.~Mainieri\inst{7}
\and
A.~Renzini\inst{8}
\and  
A.~Bongiorno\inst{9}
\and
G.~Coppa\inst{5,10}
\and
S.~de~la~Torre\inst{1,11,4}
\and
L.~de~Ravel\inst{1}
\and
P.~Franzetti\inst{4}
\and
B.Garilli\inst{4}
\and
J.-F.~Le~Borgne\inst{6}
\and
V.~Le~Brun\inst{1}
\and
M.~Mignoli\inst{5}
\and
R.~Pell\`o\inst{6}
\and
Y.~Peng\inst{3}
\and
E.~Perez-Montero\inst{6,12}
\and
E.~Ricciardelli\inst{13}
\and
J.D.~Silverman\inst{3}
\and
M.~Tanaka\inst{7}
\and  
A.M.~Koekemoer\inst{14}
\and
N.~Scoville\inst{15}
 \and  
U.~Abbas\inst{1,16}
\and
D.~Bottini\inst{4}
\and
A.~Cappi\inst{5}
\and
P.~Cassata\inst{1,17}
\and
A.~Cimatti\inst{10}
\and 
L.~Guzzo\inst{11}
\and 
A.~Leauthaud\inst{18}
\and 
D.~Maccagni\inst{4}
\and 
C.~Marinoni\inst{19}
\and 
H.J.~McCracken\inst{20}
\and 
P.~Memeo\inst{4}
\and 
B.~Meneux\inst{9,21}
\and 
C.~Porciani\inst{22}
\and 
R.~Scaramella\inst{23}
}

   \offprints{O. Cucciati (olga.cucciati@oamp.fr)}

 \institute{
 {Laboratoire d'Astrophysique de Marseille, UMR 6110
CNRS-Universit\'e de Provence, 38 rue Frederic Joliot-Curie, F-13388
Marseille Cedex 13, France}
\and
  {INAF-Osservatorio Astronomico di Brera - Via Brera 28, I-20121 Milano, Italy}
\and
 {Institute of Astronomy, ETH Z\"urich, CH-8093, Z\"urich, Switzerland}
\and
 {INAF - IASF, via Bassini 15, I-20133,  Milano, Italy}
\and
 {INAF - Osservatorio Astronomico di Bologna, via Ranzani 1, I-40127 Bologna, Italy}
\and
 {Laboratoire d'Astrophysique de Toulouse-Tarbes, Universit\'{e} de Toulouse, CNRS, 14 avenue Edouard Belin, F-31400 Toulouse, France}
\and
 {European Southern Observatory, Karl-Schwarzschild-Strasse 2, Garching, D-85748, Germany}
\and
 {INAF - Osservatorio Astronomico di Padova, Padova, Italy}
\and
 {Max-Planck-Institut f\"ur extraterrestrische Physik, D-84571 Garching, Germany}
\and
 {Dipartimento di Astronomia, Universit\'a di Bologna, via Ranzani 1, I-40127, Bologna, Italy}
\and
 {INAF-Osservatorio Astronomico di Brera - Via Bianchi 46, I-23807 Merate (LC), Italy}
\and
 {Instituto de Astrof\'\i sica de Andaluc\'\i a - CSIC.
Apdo. de correos 3004. 18080. Granada (Spain)           }
\and
 {Dipartimento di Astronomia, Universit\'a di Padova, vicolo Osservatorio 3, I-35122 Padova, Italy}
\and
 {Space Telescope Science Institute, 3700 San Martin Drive, Baltimore, MD 21218, USA}
\and
 {California Institute of Technology, MS 105-24, Pasadena, CA 91125, USA.}
\and
 {INAF - Osservatorio Astronomico di Torino, Strada Osservatorio 20, I-10025 Pino Torinese, Italy}
\and
 {Dept. of Astronomy, University of Massachusetts at Amherst}
\and
 {Physics Division, MS 50 R5004, Lawrence Berkeley National Laboratory, 1 Cyclotron Rd., Berkeley, CA 94720, USA}
\and
 {Centre de Physique Th\'eorique, UMR 6207 CNRS-Universit\'e de Provence, F-13288, Marseille, France}
\and
 {Institut d'Astrophysique de Paris, UMR 7095 CNRS, Universit\'e Pierre et Marie Curie, 98 bis Boulevard Arago, F-75014 Paris, France.}
\and 
 {Universitats-Sternwarte, Scheinerstrasse 1, D-81679 Muenchen, Germany}
\and
 {Argelander Institut f\"ur Astronomie, Auf dem H\"ugel 71,D-53121 Bonn, Germany}
\and
 {INAF, Osservatorio di Roma, Monteporzio Catone (RM), Italy}
}

   \date{Received -; accepted -}


  \abstract
   {}
   {With the first $\sim$10000 spectra of the flux limited zCOSMOS
   sample ($I_{AB}\leq 22.5$) we want to study the evolution of
   environmental effects on galaxy properties since $z \sim 1.0$, and
   to disentangle the dependence among galaxy colour, stellar mass and
   local density.}
   {  We use our previously derived 3D local density contrast
   $\delta$, computed with the $5^{th}$ nearest neighbour approach, to
   study the evolution with $z$ of the environmental effects on galaxy
   U-B colour, D4000$\AA$ break and [OII]$\lambda$3727 equivalent
   width (\emph{EW}[OII]). We also analyze the implications due to the
   use of different galaxy selections, using luminosity or stellar
   mass, and we disentangle the relations among colour, stellar mass
   and $\delta$ studying the colour-density relation in narrow mass
   bins. }
   {We confirm that within a luminosity-limited sample ($M_B
   \leq -20.5-z$) the fraction of red ($U-B \geq 1$) galaxies depends
   on $\delta$ at least up to $z\sim1$, with red galaxies residing
   mainly in high densities. This trend becomes weaker for increasing
   redshifts, and it is mirrored by the behaviour of the fraction of
   galaxies with D4000$\AA$ break $\geq 1.4$.  We also find that up to
   $z\sim1$ the fraction of galaxies with $\log(EW[OII]) \geq 1.15$ is
   higher for lower $\delta$, and also this dependence weakens for
   increasing $z$. Given the triple dependence among galaxy colours,
   stellar mass and $\delta$, the colour-$\delta$ relation that we
   find in the luminosity-selected sample can be due to the broad
   range of stellar masses embedded in the sample. Thus, we study the
   colour-$\delta$ relation in narrow mass bins within mass complete
   subsamples, defining red galaxies with a colour threshold roughly
   parallel to the red sequence in the colour-mass plane.  We find
   that once mass is fixed the colour-$\delta$ relation is globally
   flat up to $z\sim 1$ for galaxies with $\log(M/M_{\odot}) \gtrsim
   10.7$. This means that for these masses any colour-$\delta$
   relation found within a luminosity-selected sample is the result of
   the combined colour-mass and mass-$\delta$ relations. On the
   contrary, even at fixed mass we observe that within $0.1 \leq z
   \leq 0.5$ the fraction of red galaxies with $\log(M/M_{\odot})
   \lesssim 10.7$ depends on $\delta$. For these mass and redshift
   ranges, environment affects directly also galaxy colours.}
   {We suggest a scenario in which the colour depends primarily
   on stellar mass, but for an intermediate mass regime ($10.2
   \lesssim \log(M/M_{\odot}) \lesssim 10.7$) the local density
   modulates this dependence. These relatively low mass galaxies
   formed more recently, in an epoch when more evolved structures were
   already in place, and their longer SFH allowed environment-driven
   physical processes to operate during longer periods of time.}


   \maketitle
%


\section{Introduction}

Galaxies can be considered the building blocks of the universe, and
nevertheless we are far from a complete understanding of their
formation and evolution. Although theoretical scenarios can help us to
address these open issues, only through observations we can robustly
shape a coherent picture, following backward galaxy lifetimes from
what we see in the local universe up to their undisclosed origins.

The first observations of nearby galaxies led to their morphological
classification (\citealp{hubble26}, but see also \citealp{sandage75}).
In more recent years, several internal physical galaxy properties have
been used in order to classify galaxies. The inter-correlations and
scaling laws that were found among these properties and also with
morphology (to give only a few examples, the `Fundamental Plane',
\citealp{djorgovski87,dressler87}; the `Photometric Cube',
\citealp{scodeggio2002_cube}; the colour-magnitude relation,
\citealp{visv_sandage1977_CMR,tully1982_colmag,gavazzi96}) have been
confirmed in the last years using the data provided by large local
galaxy surveys like the Sloan Digital Sky Survey
(SDSS,\citealp{york2000_SDSS}) and the Two degree Field Galaxy
Redshift Survey (2dFGRS, \citealp{colless2001_2dF}), with the studies
by \cite{kauffmann2004}, \cite{baldry2004_bimodality},
\cite{balogh2004b} and several others. These findings suggested a
broad bimodal galaxy classification: 1) the red, passive galaxies,
with elliptical/spheroidal shapes and composed by an old (on the mean)
population of stars, moving mostly on disordered radial orbits; 2) the
blue, active galaxies, disk-dominated, where stars (of relatively
young age) are concentrated in circular orbits.

The close relationship between these galaxy types and local
environment was primarily uncovered via the study of nearby clusters.
The so-called morphology-density relation, as first quantified by
\cite{oemler1974} and \cite{dressler1980}, holds that star-forming,
disk-dominated galaxies tend to reside in regions of lower galaxy
density, such as cluster outskirts and the field, while red,
elliptical galaxies are found preferentially in the higher density
regions, like cluster cores. Recent work based on the 2dFGRS and the
SDSS has established that the connections between local environment
and morphology holds not only in clusters, but extend continuously
over the full range of local densities, from the centers of clusters
out into the field population. Moreover, not only morphology, but
several other galaxy properties correlate with environment. On
average, high density regions are populated mostly by redder, brighter
and less star forming galaxies, while the opposite is true for the low
density environment (\eg \citealp{balogh2004b, kauffmann2004,
tanaka2004, blanton2005, hogg2004}).  These studies also indicated
that colour is the property most tightly related to environment.

With the recent advent of large deep galaxy surveys, it has been found
that the galaxy bimodal distribution is not characteristic of the
local universe only. On the one side, early type galaxies are found to be
the predominant population among red and luminous galaxies at all
redshift up to $z\sim1.3$ (see \eg
\citealp{Bell2004,menanteau2006_ACS}). On the other side, it has been
observed that the bimodal colour distribution holds even up to
$z\sim2$, although with a significant contribution of dusty star
forming galaxies to the red population (\eg,
\citealp{willmer2006,franzetti2007_bimodality}).

Also the environment has been shown to extend its influence on galaxy
properties at these high redshifts. In the last decade studies of
increasingly distant clusters (to $z \sim 1$) have grown in number,
thanks to high-resolution imaging and spectroscopic data. These works
have shown, for example, that the fraction of early type galaxies
decreases from the inner part of clusters to their outskirts (\eg,
\citealp{treu2003}), and the fraction of star forming galaxies is
higher in the field than in any other environment (\eg,
\citealp{poggianti2006}). Moreover, it has been shown that, at least
up to $z\sim1$, the fraction of blue
galaxies anti-correlates with group richness \citep{iovino2010_groups,cucciati2009_groups},
while the fraction of early-type galaxies 
correlates with it \citep{kovac2010_groups}.  Still, comparisons of
local results with studies of high-redshift clusters have pointed
towards significant evolution in the relationship between galaxy
properties and local environment from $z \sim 1$ to the present. For
example, the morphology-density relation for galaxies in clusters and
groups evolves strongly with increasing redshift \citep{dressler1997},
and there is evidence that the evolution in intermediate density
environments has occurred more recently than in high density
environments \citep{smith05}. It also appears that the morphological
mix of cluster early type galaxies (the relative fraction of
lenticular and elliptical galaxies) has changed with time
\citep{fasano2000, postman05}.

Nevertheless, with these high redshift cluster studies it is possible
to divide galaxies only into broad environment classifications (such
as field, group and cluster populations).  Furthermore, clusters
include only a relatively small fraction (by number and an even
smaller one by volume) of the total galaxy population at any epoch,
and therefore they offer only a limited vision on galaxy evolution.
The new generation of large and deep redshift surveys very recently
allowed us to extend to high redshift the investigation of
environmental effects on several galaxy properties within the full
range of local densities, from voids and very low densities up to the
higher density peaks in the center of clusters. These high redshift
studies could help us resolve the still open debate about the origin
of environmental effects on galaxy properties. The question is whether
these properties were imprinted upon the galaxy population at their
formation epoch (the so called `nature' hypothesis), or whether they
are a result of environment-driven evolution, \ie the end products of
processes that have operated over a Hubble time (the so called
`nurture' scenario).

Many physical mechanisms that could be responsible for this
correlation between galaxy properties and environment have been
proposed, but up to now there is no a clear understanding of their
real role on the observed trends. From preliminary studies of local
density and group membership at $z \sim 1$ that have been developed
independently and nearly at the same time in these past few years by
the VIMOS-VLT Deep Survey (VVDS, \citealp{lefevre2005a}) and by the
DEEP2 Galaxy Redshift Survey \citep{Davis2003}, it seems at least that
both a biased galaxy formation (galaxies formed earlier in high
matter-density peaks, giving galaxies different imprinting of initial
conditions, \citealp{marinoni2005}) and a complex evolution of star
formation quenching (for example the weakening of the colour-density
relation at higher redshift, \citealp{cucciati2006} and
\citealp{cooper2007col}, with possibly its reversal at $z\sim 1.5$,
\citealp{cucciati2006}, but see also \citealp{elbaz2007_SFR} and
\citealp{cooper2008sfr}) concurred to build up the observed relations
between environment and galaxy properties.  We will discuss more
deeply high redshift results in the following of this paper.

A detailed view of the environmental effects on galaxy properties is
still missing, however.  The scale length at which
environment affects the most galaxy properties has been studied in the
local universe, and it seems that local density acts on colour,
$H\alpha$ equivalent width and D4000$\AA$ break on a scale of $\sim 1$
Mpc $h^{-1}$ (\citealp{kauffmann2004}, but also \citealp{blanton2006}
for scales also $\sim 0.5$ Mpc $h^{-1}$), and what it is observed on
larger scales is only a mirror of what really happens on smaller
distances. Nevertheless, these results are related only intuitively at
the typical scales of cluster halos, and a precise quantification of
these observed effects has still to be done. Moreover, as it
is known that several properties are inter-related and they also are
related with environment, one would like to understand whether the
local density affects only a property, that then drives the dependence
of the others, or more than one.  For example, we already know that
mass is responsible for shaping galaxy star formation history
\citep{gavazzi1996_massSFH,scodeggio2002_cube,kauffmann2003_massSFH},
and moreover mass does depend on environment, as it as been shown at
both local and high ($z\sim1$) redshift
\citep{kauffmann2004,bundy2006,scodeggio2009_VVDSmass,bolzonella2008_MFenv}. 
Should we consider the galaxy stellar mass as the only property
directly affected by the environment, this meaning that the other
properties depend on local density only through their dependence on
mass?  The answer seems not to be unique, as at $z\sim0.1$ the galaxy
colour, D4000$\AA$ break and sSFR are found to depend on environment
even when mass is fixed \citep{kauffmann2004,baldry2006_mass}, but no
colour-density relation has been found in the range $0.2\leq z \leq
1.4$ in mass bins within VVDS data \citep{scodeggio2009_VVDSmass}.

In this frame of open issues, we use the first $\sim 10000$ zCOSMOS
spectra to inspect more deeply the environmental effects up to
$z\sim1$. With its pure flux limit ($I_{AB}<22.5$) and small redshift
measurement error ($\sim 100$km/s), this galaxy sample allowed us to
reliably reconstruct the local density field on small scales up to
$z\sim1$, as described in \cite{kovac2010_density}. Moreover the
quality of the spectra and the multi-wavelength coverage provide us
with high quality measurement of several galaxy properties.  The paper
is organized as follows. In Sections \ref{real_data} and
\ref{mass_vs_counts} we describe the galaxy sample, the galaxy
properties we are interested in, how the density contrast has been
measured and all the advantages and drawbacks of this measurement.  In
Section \ref{red_lum_dependence} we study the redshift evolution of
the colour-density relation, together with the dependence on
environment of other galaxy properties (D4000 $\AA$ break,
[OII]$\lambda$3727 equivalent width). In Section
\ref{mass_segr_section} we show how stellar mass depends on local
density, and in Section \ref{mass_colour} we disentangle the
correlation among colour, stellar mass and environment.  Finally in
\ref{scales_section} we discuss the scale dependence of environmental
effects.  Sections \ref{comp_literature}, \ref{discussion} and
\ref{conclusions} are devoted to the discussion of our results and to
the conclusions.  Throughout this paper we use the cosmological
parameters $\Omega_m=0.25$, $\Omega_{\Lambda}=0.75$, $H_0=70$ km
s$^{-1}$ Mpc$^{-1}$ and $H_0 = 100~h$. Magnitudes are expressed 
in the AB system, and absolute magnitudes have $H_0=70$ incorporated. 
Moreover, ``mass'' always means ``stellar mass'',
unless differently stated.

Parallel analyses about galaxy properties and environment have been
carried out within the same data set in other works. Using the density
field presented in \cite{kovac2010_density},
\cite{bolzonella2008_MFenv} and \cite{zucca2009_LF} study the Galaxy
Mass Function and Galaxy Luminosity Function as a function of
environment, \cite{tasca2009} show how the local density affects
galaxy morphology and finally \cite{caputi2009_24menv} and
\cite{silverman2008_AGNenv} study the environment surrounding specific
populations (24 {$\mu m$} galaxies and AGN respectively). Moreover,
exploiting the group catalog presented in \cite{knobel2009_groups},
\cite{iovino2010_groups} and \cite{kovac2010_groups} study galaxy
colour and morphology for group and isolated galaxies.  Finally, 
\cite{peng2010_picture} draw a global picture on the role of
mass and environment in influencing galaxy evolution.


\section{DATA}\label{real_data}

\subsection{The zCOSMOS ``10k-sample''}

zCOSMOS \citep{lilly2007_zCOSMOS} is an on-going large spectroscopic
survey, conceived with the main goal of studying high-redshift galaxy
environments from very small scales (galaxy groups scales) up to
larger-scale structures. Two main projects are part of the zCOSMOS
survey: a \emph{bright} survey, aimed to measure spectroscopic
redshifts for $\sim 20000$ galaxies with $I_{AB}\leq22.5$ in the $\sim
1.7$ deg$^2$ COSMOS field \citep{scoville2007_COSMOS}, and a
\emph{deep} survey, with the goal of acquiring $\sim 10000$ spectra
for a sample of colour-selected galaxies residing in the redshift
range $1.4<z<3.0$, in the central 1 deg$^2$ of the COSMOS field. The
results described in this paper are based on the first $\sim10000$
measured spectra in the \emph{bright} sample
\citep{lilly2009_zCOSMOS}. From now on, we will call this sample
``10k-sample''.

Details about the zCOSMOS survey can be found in
\cite{lilly2007_zCOSMOS} and \cite{lilly2009_zCOSMOS}. In summary,
spectra have been collected with the VIMOS \citep{lefevre2003_VIMOS}
multi-spectrograph, at ESO Very Large Telescope. \emph{Bright} sample
observations have been performed using the Red Medium-Resolution grism
($R\sim600$, spectral range 5550-9650 $\AA$).  The final \emph{bright}
sample of $\sim 20000$ will have an estimated sampling rate of
$\sim60-70 $\% homogeneous over almost the entire field, while the
10k-sample has a less homogeneous and lower sampling rate, on average
of $\sim33$\%, and it actually covers an area of $\sim 1.4$ deg$^2$.

For the data reduction we refer the reader to
\cite{lilly2009_zCOSMOS}. Here it is worth noticing that thanks to
repeated observations carried out on a subsample of galaxies, a
measurement redshift error of $\sim 100$ km/s has been estimated.  For
this paper, a subsample of galaxies with an overall 98.4\% reliability
rate in the redshift measurement have been used. This leave us with a
sample of $\sim7800$ galaxies in the redshift range $0.1 \leq z \leq
1.0$.

Finally, the COSMOS field has been covered by multi-wavelength
imaging. Optical and Near-IR data are described in
\cite{capak2007_photcat}. See references therein for details on
observations in other wavelengths (XMM-Newton, GALEX, VLA) and
\cite{koekemoer07_HST} for the HST ACS imaging data.

\subsection{Galaxy properties}\label{gal_prop}

In this paper we analyze environmental effects on galaxy properties
such as absolute magnitude, stellar mass and spectral indexes.

Rest frame absolute magnitudes have been computed from the best
fitting template normalized to each galaxy photometry and redshift,
using the code \emph{ZEBRA} \citep{feldmann2006_ZEBRA}. The
template set is composed by six observed galaxy spectra (four from
\citealp{CWW1980} and two from \citealp{kinney1996}).  We refer the
reader to \cite{oesch2008_magabs} for more details. We tested the
robustness of our results using also absolute magnitudes derived with
the method described in \cite{ilbert2005}, and with two different
template sets: the library from \cite{BC03}, hereafter BC03, and
stellar population models generated using the PEGASE2 population
synthesis code \citep{pegase1997}. We obtained consistent results,
within error bars, for different choices of template set and/or code.

Galaxy stellar masses have been computed using a SED fitting
technique, with the \emph{Hyperzmass} code (a modified version of
\emph{Hyperz}, see \citealp{bolzonella2000_hyperz}). Different SED
libraries have been used: the widely-used BC03 library, the
\citet[hereafter M05]{maraston2005} library and Charlot \& Bruzual
(2007, hereafter CB07, private communication) library.  The imposed
Star Formation Histories are 10 exponentially declining SFH with
\emph{e-}folding times from 0.1 to 30 Gyr, plus a model of constant
star formation. In this paper we use masses derived with BC03
libraries, a Chabrier Initial Mass Function \citep{chabrier2003_IMF},
and dust extinction modeled with the Calzetti's law
\citep{calzetti2000_dust}. For further details and comparisons among
different libraries see \cite{pozzetti2008_MF}.  It is worth noticing
that results obtained in this paper regarding stellar masses are
robust using masses derived with different libraries.

In this work me make use of two spectral features: the equivalent
width of the [OII]$\lambda$3727 doublet (\emph{EW}[OII] from now on)
and the amplitude of the 4000$\AA$ break in its narrow definition
($D_{n}$4000 from now on), as introduced by \cite{balogh1999}. These
are computed by an adapted version of the \emph{platefit} code
(\citealp{lamareille2006_platefit}, see also \citealp{tremonti2004_platefit}
and \citealp{brinchmann2004_platefit} for the original pipeline). 
This code fits the stellar continuum
and absorption lines making use of the STELIB library
\citep{leborgne2003_STELIB} of stellar spectra and of the GALAXEV
\citep{BC03} stellar population synthesis models. We eliminate from
our sample spectral measurements affected either by large
uncertainties in the wavelength calibration, by the superposition with
strong night sky lines or by stellar continuum subtraction problems.
Finally, we consider only spectral line measurements with a line
significance (ratio of the maximum flux of a line to the {\it
rms} of the continuum around it) higher than 1.15. This threshold
allows us to exclude $\sim$85\% of fake detections, and still to
include the majority of real emission lines measurements.

For $D_{n}$4000 and [OII]$\lambda$3727 related measurements, we
consider for our analysis the redshift range $0.48 \leq z \leq 1.0$,
as the [OII]$\lambda$3727 line enters in the spectral range of the
adopted grism only at $z=0.48$ (the D4000$\AA$ break enters at a
relatively smaller $z$, but we want to consider the same $z$ range for
the two indicators). In this redshift range, $D_{n}$4000 has been
reliably computed for $\sim4000$ galaxies, and we have a successful
measurement of [OII]$\lambda$3727 equivalent width for $\sim2500$
galaxies.


\begin{figure} 
\begin{center}
\includegraphics[width=9cm]{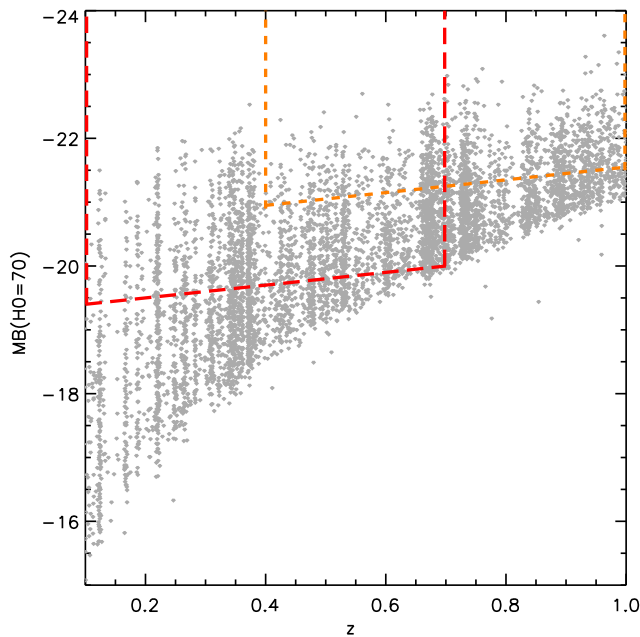}
\caption{B-band rest frame absolute magnitudes as a function of
redshift for galaxies of the 10k-sample residing in the central area
of the zCOSMOS field, in the redshift range considered in this
paper. Two time-evolving luminosity volume limits are shown with
dashed lines: the red long-dashed line represents the fainter limit
($M_B = -19.3 -z$), for which we are complete up to $z=0.7$, and the
orange short-dashed line is the limit $M_B = -20.5 -z$, for which we
are complete up to $z=1.0$ but for which we have enough galaxies only
for $z \geq 0.4$. } 
\label{spanh} 
\end{center} 
\end{figure}


\begin{figure} \begin{center}
\includegraphics[width=8cm]{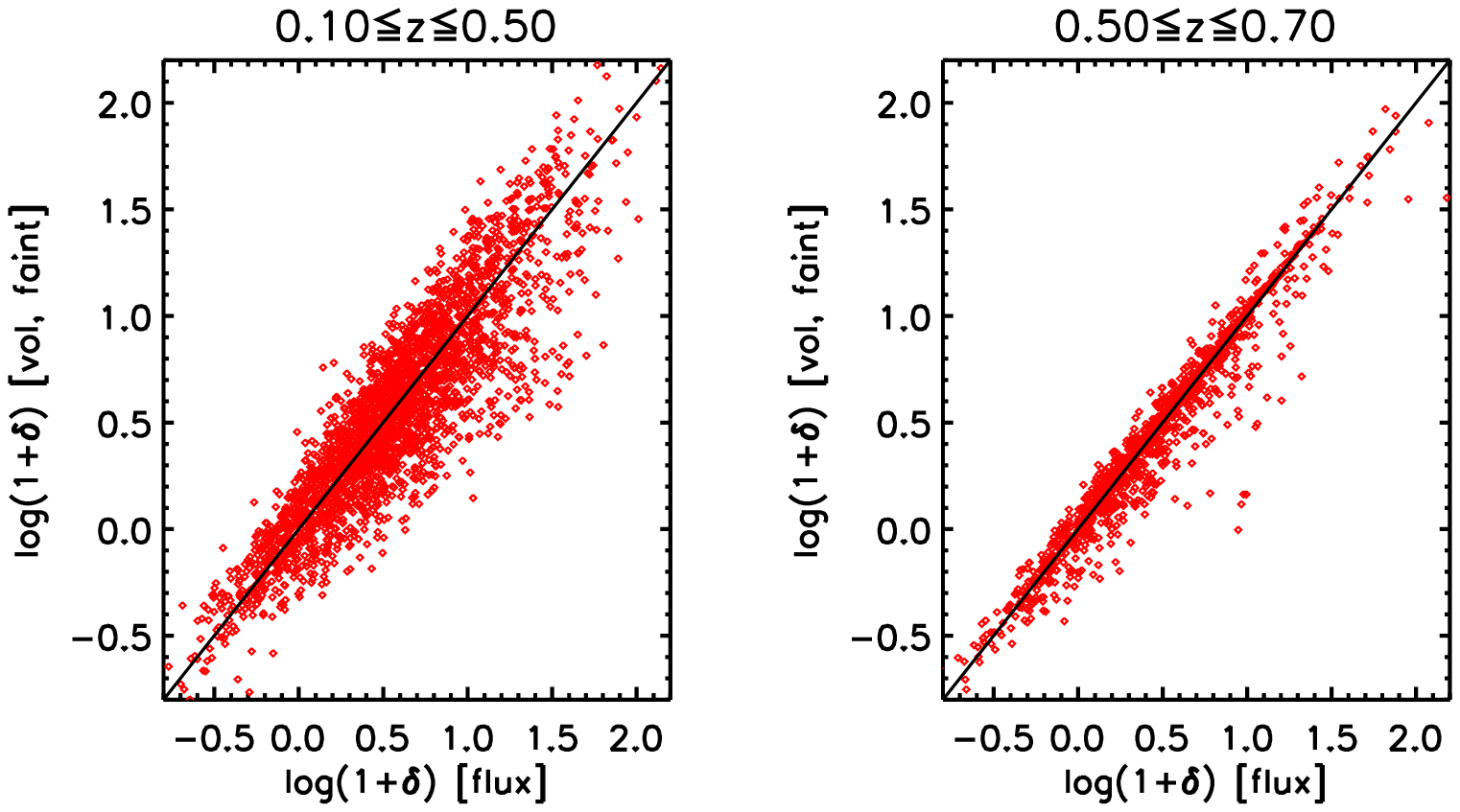}
\includegraphics[width=8cm]{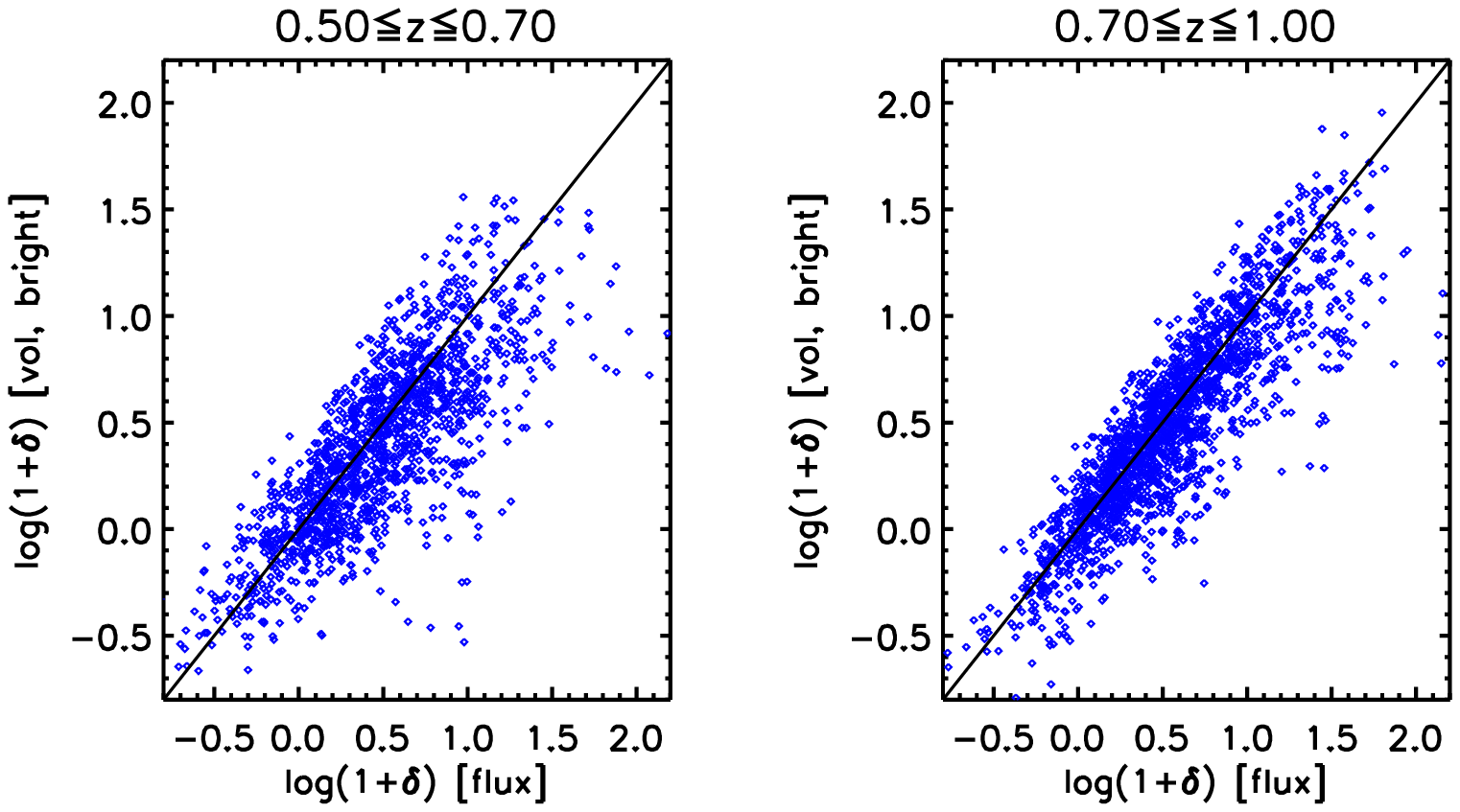}
\caption{Comparison of local density computed with volume limited
tracers (`vol', $y$ axis) and flux limited tracers (`flux', $x$ axis). The density
computed with the faint and bright volume limited tracers is shown in
the top and bottom panels, respectively, in each case for two redshift
bins as indicated on the top of each plot.} 
\label{comp_den} 
\end{center} 
\end{figure}

\subsection{Density}\label{density}

The detailed description of the environment parametrization within
the zCOSMOS 10k-sample is given in \cite{kovac2010_density}. Here we
give only a general overview.

The environment surrounding a given galaxy at comoving position $\vec
r$ has been characterized by the dimensionless density contrast
$\delta(\vec r) = [\rho(\vec r) - \rho_{m}(\vec r)]/\rho_{m}(\vec r)$,
where density is smoothed with a given filter. In this formula,
$\rho_{m}(\vec r)$ is the mean density at the redshift that
corresponds to the comoving position $\vec r$, while $\rho(\vec r)$ is
the local density around the given galaxy.

We remark that for the environment parametrization not only the
10k-sample galaxies has been used, but also the remaining $\sim 30000$
galaxies in the full zCOSMOS photometric catalogue, exploiting their
photometric redshifts. Moreover, boundary corrections have been
applied to density estimates, to take into account the fact that for
galaxies that reside too close to the field edges the filter window
will fall out of the survey field. Finally, the robustness of this
density reconstruction scheme has been tested with simulated galaxy
catalogues.

We refer the reader to \cite{kovac2010_density} for an overview
of the properties of the density field reconstructed in the zCOSMOS
field. In particular, their figures 22, 23, and 24 show a direct
comparison of the density field and the distribution of galaxy groups
\citep{finoguenov2007_clusters,knobel2009_groups}.  Another analysis
of the large scale structures in the same field is shown in
\cite{scoville2007_lss}. For example, they remark that the
supermassive structure at $z \sim 0.74$ seems to be equivalent, in
terms of mass, to the Coma cluster.

Given the lower and less homogeneous sampling rate near the field
boundaries, in this work we limit ourselves to the central area of the
zCOSMOS field, enclosed within the rectangular area with vertices
$ra_{min}=$149.55, $ra_{max}=$150.42, $dec_{min}=$1.75,
$dec_{max}$=2.7.

\subsection{Choosing the more suitable density computation
approach}\label{density_approach}

The local density $\rho(\vec r)$ has been computed with a variety of
methods, including different filters, different tracer galaxy
populations and different weighting schemes for the tracer galaxies.
While \cite{kovac2010_density} discuss all these possibilities, we
discuss here only the optimal choice adopted for this work.

The different methods used to estimate the density field produce
environment parametrizations with significantly different statistical
properties (for example the measurement error, or the homogeneity
across the redshift range explored), and it is important to choose the
density estimator best suited for any specific scientific analysis.

For our study, we decided to give priority to the density computed on
as small a scale as possible. Tests performed using mock catalogues
\citep{kovac2010_density} have shown that with the zCOSMOS 10k-sample
we can safely reconstruct the galaxy local environment up to $z\sim1$
using cylindrical filters with half depth of 1000 km/s, and either a
fixed radius $R \geq $  3 h$^{-1}$Mpc, or a variable
radius equal to the projected distance of at least the 5$^{th}$
nearest neighbour (`n.n.' from now on). As one can see in Fig. 6 of
\cite{kovac2010_density}, at least at the highest densities the
distance $D$ to the 5\thnn is always smaller than 3 h$^{-1}$Mpc. Thus
we will use the density computed with the 5\thnn approach. However
with this method the scale $R$ varies as a function of environment,
being larger for lower densities. This has to be taken into account in
the interpretation of our results.

Moreover, different tracer galaxy populations have been used for the
density computation: the entire flux limited sample ($I_{AB}\leq22.5$)
and two different luminosity-selected volume-limited subsamples, that
satisfy the criteria $M_B\leq -19.3-z$ and $M_B\leq -20.5-z$. The 
redshift dependence in these equations has been chosen to take into account
the effects of an average passive evolution that should include the majority of galaxy
types, and the intercepts values have been chosen in order to define
two complete samples at $z=0.7$ and $z=1.0$, respectively
\citep{kovac2010_density}.  This is clearly shown in
Fig. \ref{spanh}. The figure shows the B-band absolute magnitude
distribution as a function of redshift for the 10k-sample, in the
redshift range considered in this paper. The two time-evolving
luminosity limits are plotted as dashed lines. From the figure, it is evident
that the fainter subsample is defined to inlcude all the objects at its 
higher redshift ($z=0.7$), while the brighter subsample at $z=1$ remains about 0.5 
mag brighter than the faintest galaxies that we can see in the 10k-sample. 
This is due to the fact that our survey is flux-limited in I-band, which 
corresponds to the restframe B-band at $z\sim0.7$. Thus, our sample is 
complete in restframe B-band  at this redshift. At higher redshift, 
because of the different evolution of different galaxy populations, 
we start missing the early type galaxies at the faintest B-band magnitudes 
we can reach. The brighter magnitude limit that we used ensures us to be complete 
for all galaxy populations.

The advantage in using the flux limited tracers is that we can exploit
our entire data set, and we can measure the density contrast down to
the smallest scales adopting the 5\thnn method. There are two
drawbacks in using these tracers: the tracer population is not uniform
with redshift, as it becomes brighter for higher $z$, and the mean
distance to the 5\thnn increases with $z$.  Both these issues have to
be considered when interpreting any evolutionary environmental effect
on galaxy properties. These disadvantages are not present if we use a
luminosity-selected volume-limited sample of tracers. For these
reasons, we decided to use the density contrast computed with the
tracers limited at $M_B\leq -19.3-z$ or $M_B\leq -20.5-z$, according
to the redshift range we want to explore. The limit at $M_B\leq
-20.5-z$ at low $z$ includes too few galaxies, so these tracers can be
safely used only in the range $0.4 \leq z \leq 1.0$. In contrast, the
fainter luminosity-limited sample ($M_B\leq -19.3-z$) is complete only
up to $z=0.7$, but can be used starting from $z = 0.1$. In
Fig. \ref{comp_den} we show the comparison between the density
computed with flux limited tracers and volume limited tracers.

Finally, different weights can be assigned to the tracer galaxies. For
example, we can weight galaxies by their stellar mass or their
luminosity in a given band, or simply by their number.  As this is a
subtle issue for what concern the general definition (and thus
interpretation) of local density itself, we discuss explicitly
advantages and drawbacks of different weighting schemes in the
following Section.


\begin{figure}
\begin{center}
\includegraphics[width=4.4cm]{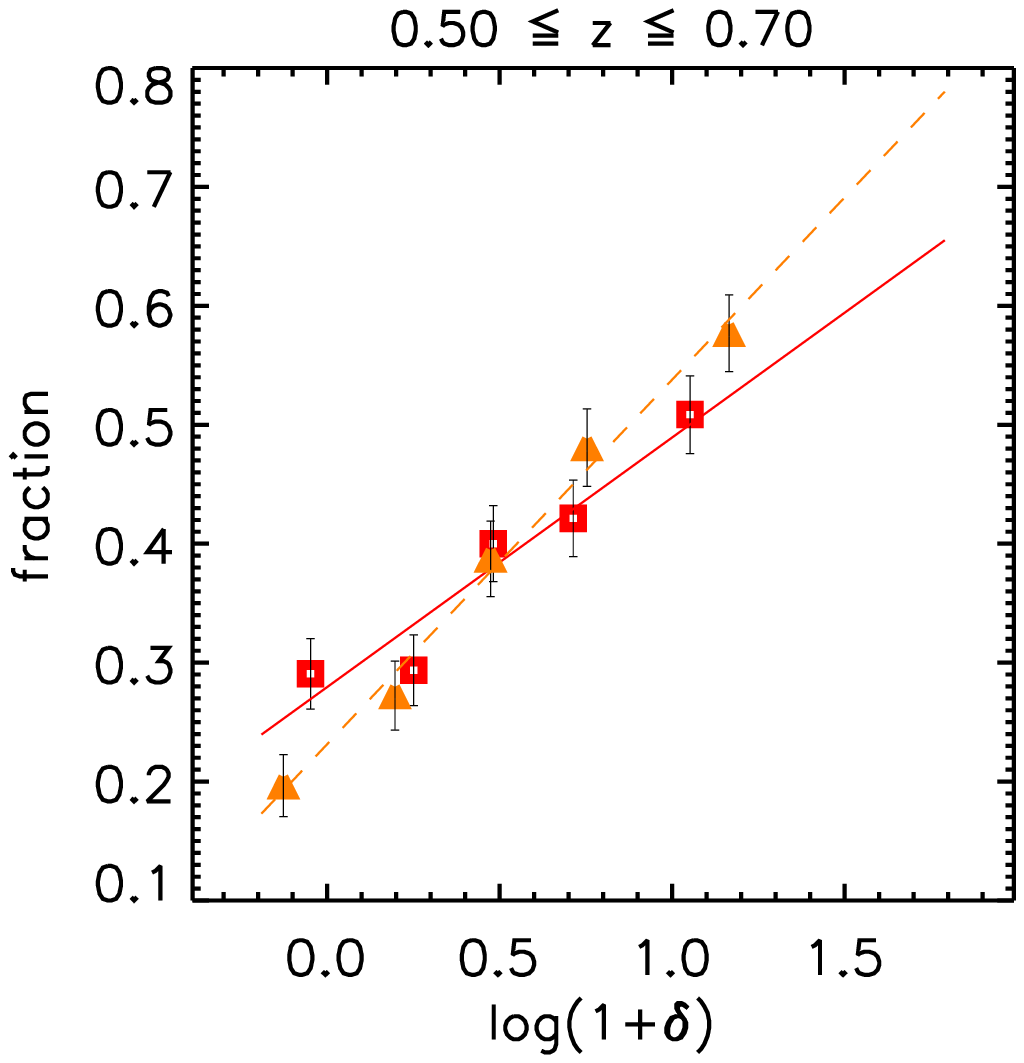}
\includegraphics[width=4.4cm]{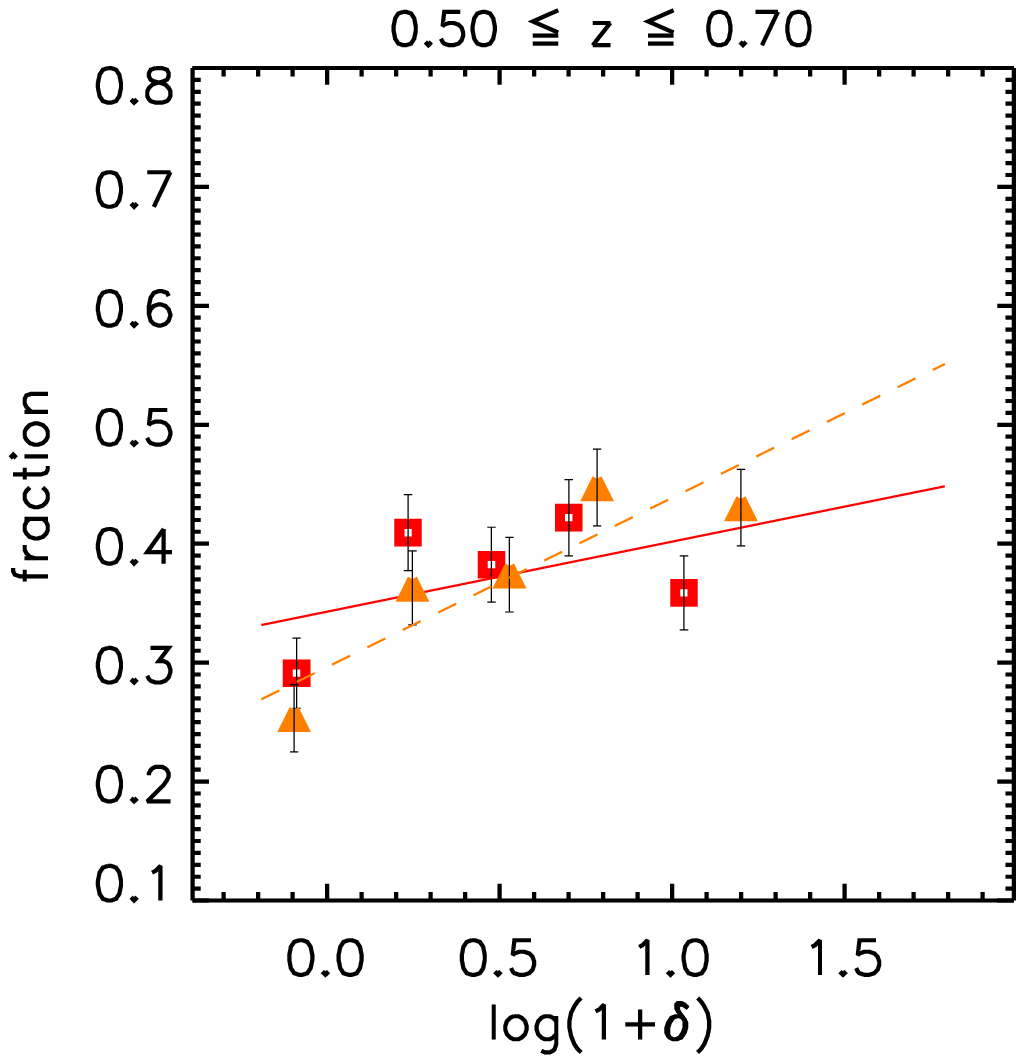}
\caption{\emph{Left panel.} Fraction of red (U-B $>$ 1.0) galaxies as
a function of the density contrast, in equipopulated density bins, for
a luminosity-selected volume-limited sample with $M_B \leq -19.3 -z$
in the redshift range $0.5\leq z \leq 0.7$.  Density is computed with
the $5^{th}$ nearest neighbour method, using volume limited tracers
($M_B \leq -19.3 -z$).  The values on the $x$ axis represent the
median density values in each density bin. Red squares are for
density based on counts (`CWD', see text), while orange triangles for
density weighted by stellar mass (`MWD'). Vertical error bars are
computed applying the binomial formula. Solid line is the linear fit
of squares, dashed line the linear fit of triangles. The slopes of the
fits with their errors are quoted in the text. \emph{Right panel.} As
in the left panel, but in this case, before computing the MWD to each
galaxy it has been assigned the entire set of properties (mass, colour
etc...) of another galaxy near in redshift, chosen randomly. This way
galaxy positions and redshift are not correlated with galaxy
properties. See the text for further details.}
\label{frac_corr_mass_vs_counts}
\end{center}
\end{figure}


\section{The mass-weighted  density and the colour-density
relation}\label{mass_vs_counts}

It has been shown that galaxy formation is `biased', in the sense that
galaxies have been formed earlier in higher matter-density peaks (see
for example \citealp{marinoni2005}). This suggests that the galaxy
total mass, if not even the (dark) matter distribution on larger
scales, is a key ingredient to understand galaxy formation and
evolution.  Moreover, we know that galaxy stellar mass correlates with
the dark matter halo mass, for both early and late type galaxies
\citep{mandelbaum2006_halo, yang2008_HOD}.  This leads to the idea
that a density estimator that takes into account the stellar mass of
the tracer galaxies could be more physically motivated than simple
galaxy counts. From now on, we will define these two different
environment parametrizations as `mass-weighted density' (MWD) and
`count-weighted density' (CWD). We also refer the reader to
\cite{kovac2010_bias}, who study in detail how the reconstructed
galaxy density field within the zCOSMOS 10k-sample relates to the
underlying matter density field.

The concern in using the MWD is the existence of a correlation between
stellar mass and other galaxy properties. The possible environmental
dependence of these properties risks to be spuriously increased by
their dependence on mass. A typical example, that moreover is the one
we are mostly interested in, is the colour-mass relation. 
The left panel of Fig. \ref{frac_corr_mass_vs_counts} shows the fraction of red
(U-B$\geq$ 1.0) galaxies as a function of the density contrast, in
equipopulated density bins, for the same luminosity-selected volume-limited
sample used for the computation of the density contrast, in the redshift range $0.5\leq z \leq
0.7$ (this cut leaves us with roughly 1200 galaxies).  
Density is computed with the $5^{th}$ nearest neighbour method, using the fainter
volume limited tracers ($M_B \leq -19.3 -z$).  Orange triangles are for MWD, while red
squares are for CWD. Vertical error bars are computed applying the
binomial formula (see Section \ref{red_lum_dependence} for
details). The dashed line is the linear fit of triangles, the solid
line the linear fit of squares. The slopes of the fits with their
errors are $0.31 +/- 0.03$ and $0.21 +/- 0.04$ for MWD and CWD
respectively. We note that the colour-density relation appears to be
stronger and more significant using MWD.

We tested the influence of the underlying colour-mass relation on this
slope change by shuffling galaxy properties, while preserving their
coordinates and redshift. The shuffle has been performed by assigning
to each galaxy the complete set of properties (mass, colour,
luminosity...) of another galaxy chosen randomly in the same narrow
redshift bin. We are aware that this choice can make the randomization
not ideal, as galaxy properties are not completely un-correlated with
galaxy 3D positions, but in this way we preserve the redshift
distribution of galaxy properties.  We then recalculated the CWD and
MWD using the real galaxy positions, but the randomly assigned
masses. This way, the CWD estimate does not change, but the 3D mass
distribution (as traced by galaxy stellar mass) is not linked any more
to galaxy position, and thus the MWD estimate is different from the
original one.

The right panel of Fig. \ref{frac_corr_mass_vs_counts} shows the
colour-density relation for the same galaxies as in the left panel ,
but in this case we used the new CWD and MWD and the `shuffled'
colours.  Symbols are the same as in the left panel. As expected,
there is no significant colour-density relation when using the CWD and the random
colours, as these colours are completely uncorrelated with the
original galaxy position. The measured slope of the linear fit is
$0.06 +/- 0.04$.  The fit is not totally flat, possibly due to the
non-optimal randomization.  On the contrary, when we use the MWD, we
do see a colour-density relation (the measured slope of the linear fit
is $0.14 +/- 0.03$). As we know that the colour is no more (or very
weakly) related to the galaxy position, but it does depend on mass,
the difference that we see in the colour-density relation when using
CWD and MWD is a spurious effect produced by the colour-mass relation.
Although the mass-weighted density may be considered a physically
motivated environment estimator, this test suggests us that it is
safer to use the density computed with pure counts. Thus we will use
the CWD in this work.

Summarizing our overall choice, in this work we will use a density
contrast estimator based on the 5\thnn method, computed with
volume-limited and `count-weighted' tracers.


\begin{figure}
\begin{center}
\includegraphics[width=9cm]{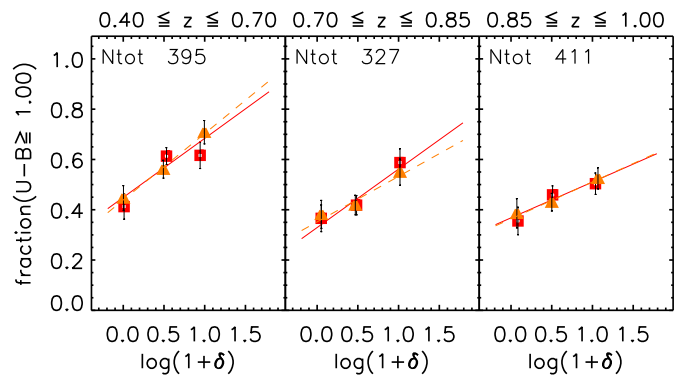}
\caption{Fraction of red galaxies as a function of the density
contrast. Galaxies are defined `red' when they have $(U-B) \geq
1.0$. Different symbols are for different density measurement methods:
red squares for bright volume limited tracers
($M_B\leq-20.5-z$), while orange triangles for flux limited
tracers.  The values on the $x$ axis represent the median density
values in each density bin. The continuous red line is linear fits of
the squares, and dashed orange line is the linear fit of the
triangles. Their slopes with associated errors can be found in Table
\ref{table_slopes}. `Ntot' is the total number of galaxies considered
in each panel. }
\label{plot_frac_col_orig}
\end{center}
\end{figure}

\begin{figure}
\begin{center}
\includegraphics[width=9cm]{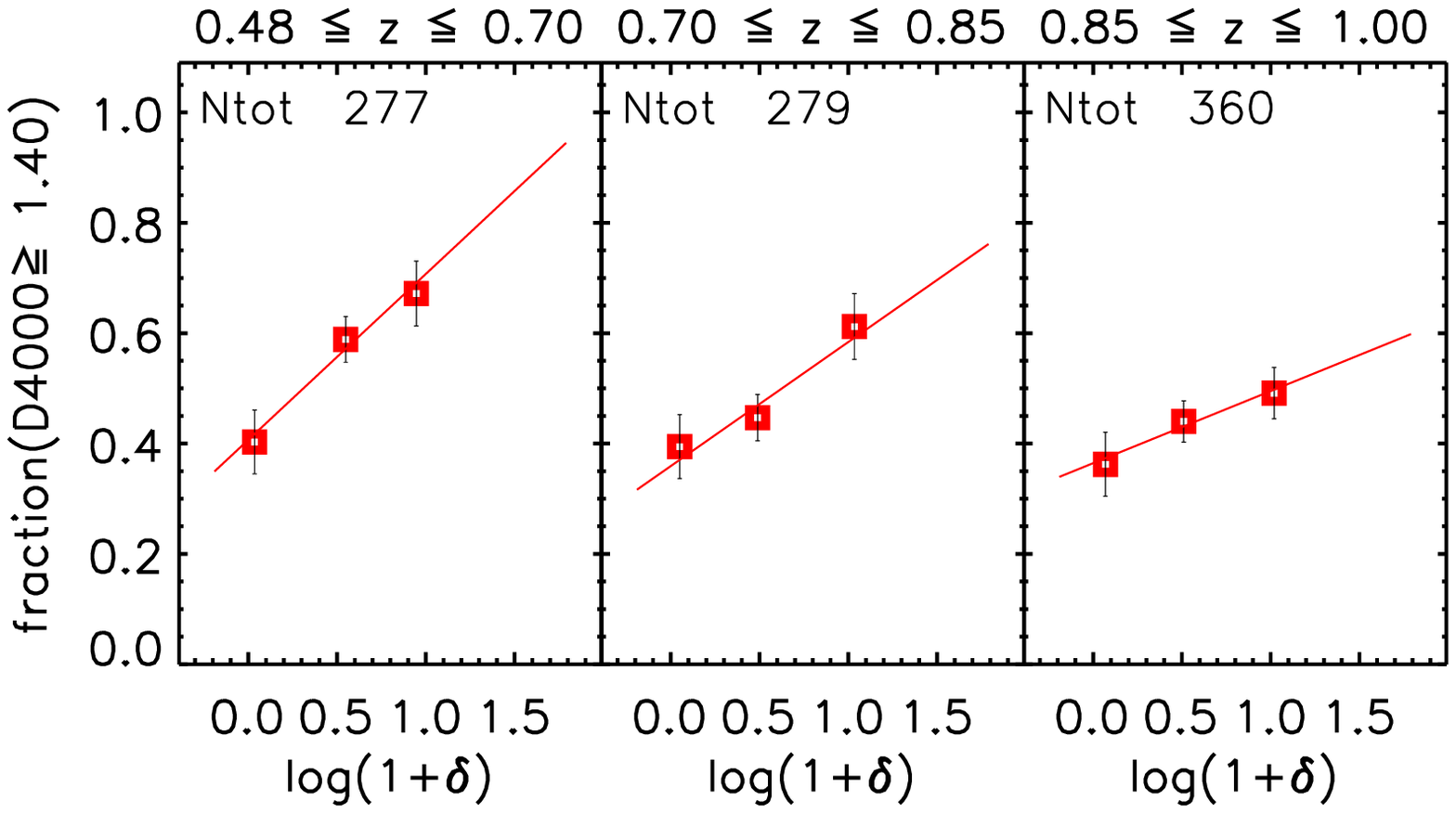}
\includegraphics[width=9cm]{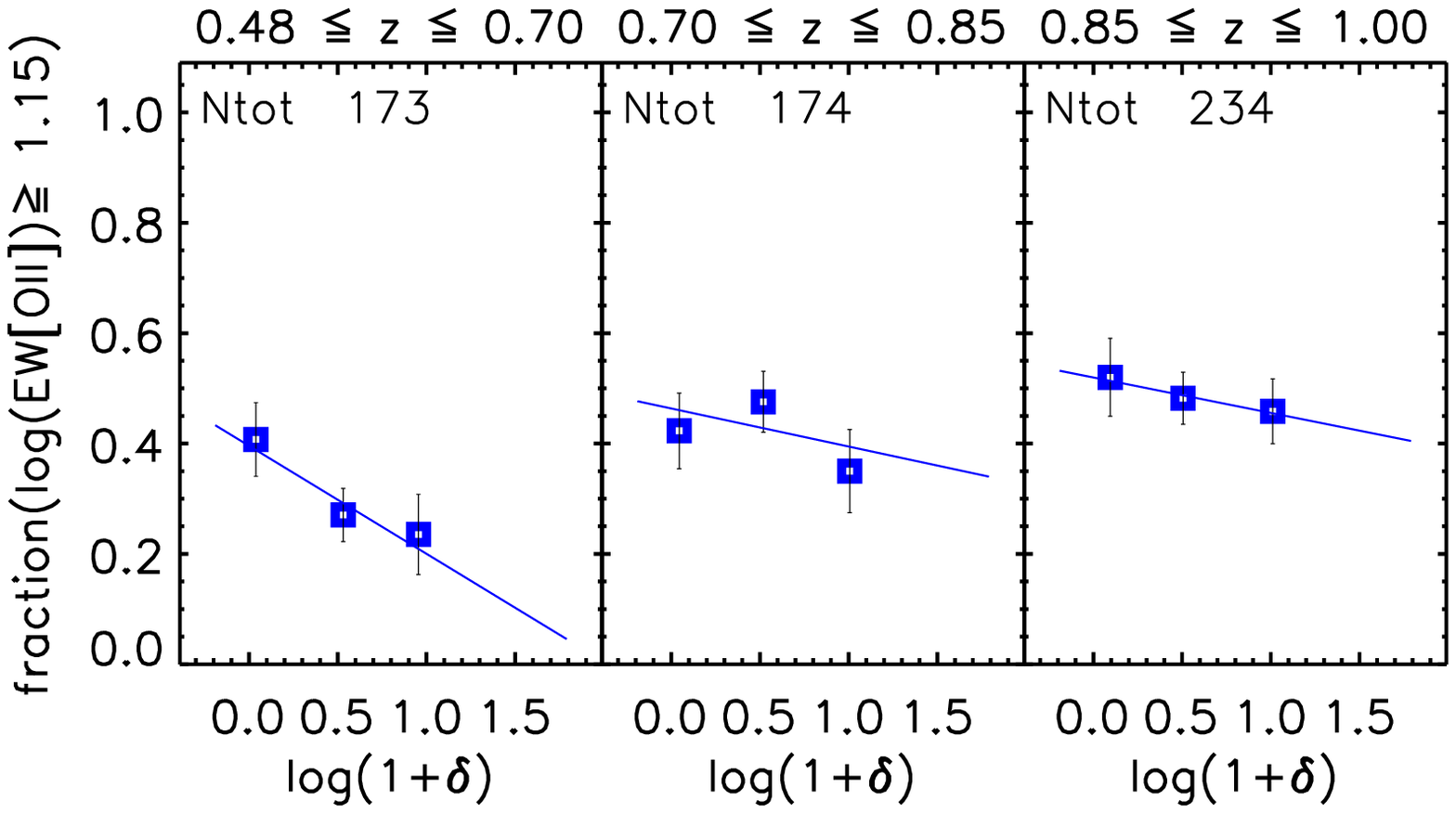}
\caption{\emph{Top.} Fraction
of `passive' galaxies as a function of the density contrast. Galaxies
are defined `passive' when they have $D_{n}4000 \geq 1.40$. Symbols
and lines have the same meaning as in the plots  of Fig. \ref{plot_frac_col_orig}.
\emph{Bottom.} Fraction of galaxies with $\log(EW[OII]) \ge 1.15$ as a function of the
density contrast. Symbols and lines as in the top panels.}
\label{plot_frac_SFR}
\end{center}
\end{figure}


\section{Redshift evolution of environmental effects}\label{red_lum_dependence}

As a first step, we want to exploit the broad redshift range covered
by the zCOSMOS 10k-sample to investigate the evolution with cosmic
time of environmental effects on galaxy properties. If we could
constrain the epoch (if any) at which environment begins playing a
role in galaxy evolution, and understand whether the properties of
galaxies residing in different environments evolve differently with
cosmic time, we could shed more light on galaxy evolution mechanisms.

In this analysis, we want to use the density computed with the
bright volume limited tracers complete up to $z=1.0$ ($M_B \leq -20.5
-z$). As discussed in Section \ref{density_approach}, with these
tracers we can reliably measure the density only for $z \geq 0.4$. So
we restrict this study to the redshift range 0.4-1.0.

The galaxy set used for this analysis is the luminosity-selected
volume-limited subsample that corresponds to the bright tracers
galaxies. As explained in Sect.\ref{density_approach}, the
dimming of 1 mag per unit redshift interval is an average evolution
that should include roughly the typical passive evolution of the
majority of galaxy types. The value of the intercept is chosen to have
a complete sample at $z=1$. This limit allows us to 
approximately follow an homogeneous and complete galaxy population
up to $z=1$. With this cut in luminosity, and considering the redshift
range $0.4\leq z \leq 1.0$, we are left with a total sample of 1133
galaxies (out of 3868 galaxies in the flux limited sample over
the same redshift range).

We analyzed the widely studied colour-density relation, focusing on
the restframe $U-B$ colour. The zCOSMOS \emph{bright} sample is selected in the
observed I-band magnitude, that corresponds to a selection in the
absolute B-band magnitude at $z\sim0.8$. This way, our sample is not
affected by any significant U-B color incompleteness up the highest
redshift we are interested in ($z \sim 1$).

The plot in Fig. \ref{plot_frac_col_orig} shows the fraction of red
galaxies (red squares) as a function of the density contrast, in three
different redshift bins. Galaxies are defined `red' when they have
$(U-B) \geq 1.0$. We choose this colour threshold as it is roughly in
the middle of the so called green valley of the bimodal colour
distribution. We keep this value fixed at all explored redshifts, as
we do not see any strong evolution of the green valley location with
$z$. The density bins are chosen so that the total sample used in this
analysis (the considered 1133 galaxies) is divided into three
equipopulated bins. This way the density bins in each panel are not
exactly equipopulated, but the environment thresholds are constant
with redshift. This implies that we are observing, at different
redshifts, comparable over-/under-densities, disregarding the average 
evolution of the density contrast. Vertical
error bars on the red fraction values are the $1\sigma$ confidence
level for the binomial statistics, computed according to the
approximation given in \cite{gehrels86_binomial}: $\sigma^2=f_r
f_b/n$, where $f_r$ is the fraction of `red' galaxies, $f_b$ is the
fraction of `blue' ones ($=1-f_r$) and $n$ is the total number of
galaxies in the density bin. When the fraction for which we want to
compute the $1\sigma$ confidence level is zero (/one), we fix its
lower (/upper) limit to zero (/one), and its upper (/lower) $1\sigma$
level at $\sigma=0.5/n$, as suggested by \cite{depropris04}.

We fit the colour-density relation with a linear fit (red solid line),
taking into account the vertical error bars of the points. We
disregarded the estimated horizontal error on the median density
value, as it is so small (on average around 0.06 dex) that the error
on the linear fit is dominated by the error on the red galaxies
fraction. In contrast, we verified that changing the number of density
bins (compatibly with the total number of galaxies) does not change
significantly the result of the fit. These considerations hold for all
our subsequent analyses.

The derived slope of
the linear fit and its error are quoted in the second column of Table
\ref{table_slopes}. The significance associated to the
colour-density trend is $\sim 3\sigma$, $\sim 3\sigma$ and $2\sigma$
from the lowest to the highest redshift bin. The colour-density
relation, present at low redshift, is still present up to
$z=1$. Moreover, there is a trend of weakening of the colour-density
relation for increasing redshift, as the slope of the fit becomes less
steep at higher redshift.

As a reference, we list in Table \ref{table_slopes} the slopes
(and their error bars) of the linear fit of the colour-density
relation that we find when using also different tracers of the local
density (volume limited and MWD, flux limited and CWD, and flux
limited and MWD, in the third, fourth and fifth column,
respectively). The reader can notice the foreseen effect of a enhanced
colour-density relation when 1) investigating the smaller scales
probed by flux limited tracers and 2) using the mass-weighted
density. The effect given by the flux limited tracers sample is
evident at least for $0.4\leq z \leq 0.7$, where the flux limited
sample is much fainter than the volume limited one (see
Fig. \ref{spanh}).

Usually, galaxy colours can be considered as a proxy for star
formation activity, with redder galaxies having less on-going star
formation. We thus analyze the dependence on environment of two other
star formation indicators: the $D_{n}$4000 and the \emph{EW}[OII]. In
particular, $D_{n}$4000 is quite well related with the $U-B$ colour,
as the U and B bands bracket the break itself. Nevertheless, the
$D_{n}$4000 is a direct measure of the stellar population, and it is
less affected by reddening or uncertainties due to the SED fitting
necessary for the colour computation (see for example in
\citealp{franzetti2007_bimodality}).

We therefore identify two logical equivalents of our `red' galaxy
sample by defining a `passive' galaxy subsample, consisting of all
galaxies having $D_{n}4000 \geq 1.4$, and an `active' galaxy subsample
including those galaxies with $\log(EW[OII]) \geq 1.15$. The reader is
referred to \cite{maier2009_SFR} and \cite{silverman2009_SFR} for more
details on the star formation activity within zCOSMOS galaxies.

These thresholds correspond roughly to the median values of the total
$D_{n}$4000 and \emph{EW}[OII] distributions. As for $U-B$ colour, the
$D_{n}$4000 threshold does not evolve significantly with redshift,
thus we used it for all the redshift bins investigated; on the
contrary, the \emph{EW}[OII] mean value increases with redshift, but
we verified that the dependence of the `active' galaxy fraction on
environment does not change significantly if we substitute the fixed
threshold with an evolving one (with the exception of a different
normalization).

In Fig. \ref{plot_frac_SFR} we show the fraction of `passive' (top
row) and `active' galaxies (bottom row) as a function of density, for
three different redshift bins (the three columns). The symbols and the
solid lines in these panels have the same meaning as those in Fig.
\ref{plot_frac_col_orig}.  Note that these features enter the observed
wavelength range at about $z=0.48$, thus the lower limit of the first
redshift bin is different from the one in
Fig. \ref{plot_frac_col_orig}.  The slopes of the linear fits,
together with their $1\sigma$ levels, are quoted in Table
\ref{table_slopes}.

Figure \ref{plot_frac_SFR} shows that the environmental effects on
$D_{n}4000$ as a function of redshift mirror those on galaxy $U-B$
colour, with comparable slopes and normalizations (inside the error
bars). As the `red' galaxies, the `passive' ones reside preferentially
in high density regions, and this $D_{n}4000$-density relation is
visible in the entire range $0.48 \leq z \leq 1$, weakening for
increasing redshift.  On the contrary, the fraction of `active'
galaxies is higher in low densities. Even if less significant than the
detection of $D_{n}4000$-density relation, we find that the
\emph{EW}[OII]-density relation holds up to $z\sim 1$, and like the
others it becomes weaker for increasing redshift. In Table
\ref{table_slopes} we quote the slopes of the linear fits of these
relations also when MWD and flux limited tracers are used to
estimate the environment. As for the colour-density relation, results
are compatible within errors irrespectively of the density estimator
used, with a general indication for stronger and more significant
trends in the MWD case. This is mainly due to the relation between
stellar mass and both $D_{n}4000$ and \emph{EW}[OII], as discussed in
Section \ref{mass_vs_counts}.  Moreover, it seems that above
$z\sim0.7$ the \emph{EW}[OII] is less sensitive to environment than
colour and $D_{n}4000$. A more detailed analysis of this issue is
deferred to future work.

We remark that we excluded from this analysis all galaxies with low
quality spectral features measurements as described in Section
\ref{gal_prop}. These galaxies are $\sim 15$\% and $\sim 45$\% of the
sample used for the colour-density analysis, considering the
$D_{n}4000$ and the \emph{EW}[OII] measurement respectively.  To be
sure that this spectral quality selection does not introduce any bias
as a function of density, we re-computed the $D_{n}4000$- and
\emph{EW}[OII]- density relations considering also these previously
excluded galaxies. We found results compatible (within errors) with
the ones shown in Fig.  \ref{plot_frac_SFR}.

Summarizing, in Fig. \ref{plot_frac_col_orig} and \ref{plot_frac_SFR}
we have shown how different galaxy properties depend on environment
within a complete luminosity-selected volume-limited sample, and how
these dependences evolve with redshift. Globally, these results
confirm already known trends: the fraction of red galaxies is higher
in high densities, and the colour-density relation becomes weaker for
increasing redshift. We have also shown that the fraction of galaxies
with high $D_{n}$4000 depends on environment in a way that is very
similar to the $U-B$ colour (as both red colour and high $D_{n}$4000
are indicators of old and passively evolving galaxies), and this is
the first time that this is shown up to $z\sim 1$.  Conversely we
found that the `active' galaxies (those with larger \emph{EW}[OII])
reside preferentially in low densities. As for colour and $D_{n}$4000,
the \emph{EW}[OII]-density relation holds up to $z\sim1$, although
with less significance, and weakens for increasing redshift, further
confirming the general trends observed with the other star formation
indicators. We refer the reader to Section \ref{comp_literature} for
the comparison of these findings with previous works, and to Section
\ref{discussion} for the discussion of their implications.

The comparative study of the environmental dependences of different
galaxy properties at different epochs is of crucial importance to shed
more light on galaxy evolution. We defer this analysis to future work,
while here from now on we will explore more in details only the
colour-density relation.

\begin{table}
\caption{ Slopes and their $1\sigma$ confidence levels of the linear fits
 shown in Fig. \ref{plot_frac_col_orig} and \ref{plot_frac_SFR}
 (second column), that is the slopes of the `property'-density
 relations when density is computed with volume limited tracers ($M_B
\leq -20.5 -z$) and
 `count-weighted' density (CWD), with `property' standing for $U-B$
 colour, $D_{n}$4000 and \emph{EW}[OII] as indicated in the first
 column. From the third to the fifth column the same slopes are
 quoted, but in these cases they are obtained with density computed
 with different tracers (volume limited and MWD, flux limited and CWD,
 flux limited and MWD, respectively).}
\label{table_slopes}
\centering
\vspace{0.2cm}
\begin{tabular}{c c c  c c }

\hline\hline

  $z$ range      & \multicolumn{2}{c}{volume limited tracers}& \multicolumn{2}{c}{flux limited tracers}\\
        & CWD & MWD  & CWD & MWD \\

\hline
{\bf U-B}& \multicolumn{4}{c}{}\\
  0.40 - 0.7         &  0.23$\pm$0.08  &  0.33$\pm$0.07 &  0.26$\pm$0.07 &  0.44$\pm$0.06 \\
  0.7 - 0.85         &  0.23$\pm$0.08  &  0.25$\pm$0.06 &  0.18$\pm$0.08 &  0.25$\pm$0.06 \\
  0.85 - 1.0         &  0.14$\pm$0.07  &  0.19$\pm$0.06 &  0.15$\pm$0.07 &  0.20$\pm$0.06 \\

\hline
{\bf $D_{n}$4000}& \multicolumn{4}{c}{}\\
  0.48 - 0.7         &  0.30$\pm$0.09  &  0.38$\pm$0.08 &  0.38$\pm$0.08 &  0.48$\pm$0.07 \\
  0.7 - 0.85         &  0.22$\pm$0.08  &  0.23$\pm$0.07 &  0.19$\pm$0.08 &  0.28$\pm$0.06 \\
  0.85 - 1.0         &  0.13$\pm$0.08  &  0.15$\pm$0.07 &  0.11$\pm$0.07 &  0.16$\pm$0.06 \\

\hline
{\bf \emph{EW}[OII]}& \multicolumn{4}{c}{}\\
  0.48 - 0.7         &  -0.20$\pm$0.11  &  -0.27$\pm$0.09 &  -0.28$\pm$0.09 &  -0.35$\pm$0.09 \\
  0.7 - 0.85         &  -0.07$\pm$0.11  &  -0.05$\pm$0.08 &  -0.06$\pm$0.10 &  -0.13$\pm$0.08 \\
  0.85 - 1.0         &  -0.06$\pm$0.09  &  -0.11$\pm$0.09 &  -0.06$\pm$0.09 &  -0.10$\pm$0.08 \\

\hline
\hline

\end{tabular}
\end{table}


\begin{figure}
\begin{center}
\includegraphics[width=9cm]{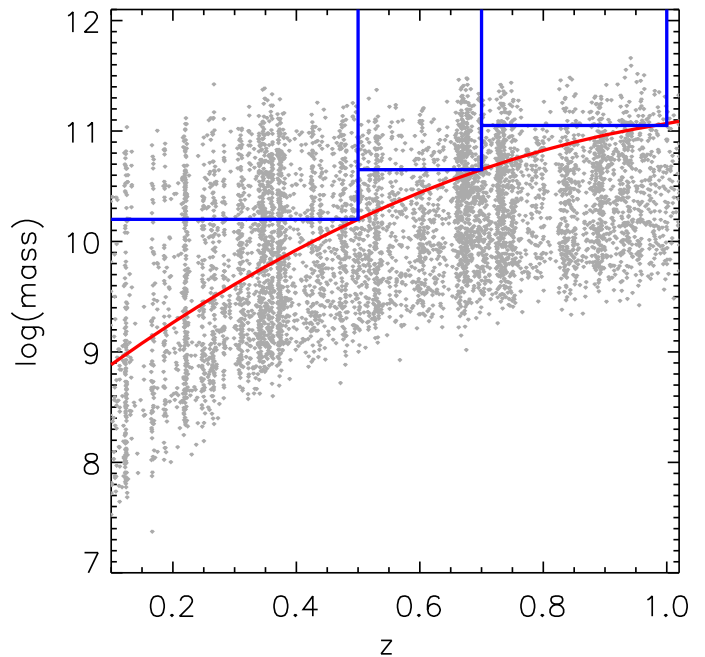}
\caption{The mass-redshift plane for galaxies of the 10k-sample
residing in the central area of the zCOSMOS field, with red curve
representing the 95\% mass completeness described in Section
\ref{mass_segr_section}. The three redshift bins considered in the
analysis and their chosen mass limits are indicated with straight
lines.}  \label{mass_limits}
\end{center}
\end{figure}


\section{Mass segregation as a function of environment}\label{mass_segr_section}

It has already been shown that galaxy stellar mass ($M$) does depend
on environment, at least weakly, at both $z\sim0.1$ and up to $z=1.4$
\citep{kauffmann2004,bundy2006,scodeggio2009_VVDSmass,bolzonella2008_MFenv}. There
is still considerable debate whether this mass-density relation is the
driver of environmental effects upon other properties of a galaxy such
as its colour and efficiency of star formation, given their known
correlations with stellar mass. For example, in the local universe
galaxy colour, $D_{n}$4000 and sSFR are found to depend on environment
even when mass is fixed \citep{kauffmann2004,baldry2006_mass}, but
this has not been found in the range $0.2\leq z \leq 1.4$ within VVDS
data \citep{scodeggio2009_VVDSmass}.  Although
\cite{scodeggio2009_VVDSmass} themselves explain the possible reasons
for this discrepancy, up to now no other study of this kind has been
performed at intermediate-high redshift. With the zCOSMOS 10k-sample
data we can now add a new insight on this issue in the range $0.1\leq
z \leq 1.0$.

The detailed analysis of the mass-density relation within the zCOSMOS
data is fully described in \cite{bolzonella2008_MFenv}, where the
Galaxy Stellar Mass Function (GSMF) is studied for different galaxy
subsamples and environments. For the sake of clarity and completeness,
in this Section we will simply present how we select a mass limited
subsample suited for our following analysis, and how the stellar mass
depends on environment in particular within this selected sample.
Then, in Section \ref{mass_colour} we will disentangle the triple
relation among colour, stellar mass and local density.

As in all flux limited samples, the range of luminosities and masses
allowed in our sample varies with redshift (see Fig. \ref{spanh} and
\ref{mass_limits}).  Given the scatter in the luminosity-mass
relation, a luminosity-selected volume-limited sample at a given
redshift will not be complete in stellar mass. As we now require a
mass-complete sample for our analysis, we can not use the
luminosity-selected volume-limited sample described in Section
\ref{red_lum_dependence}.  Since the mass-to-light ratio (M/L) is
different for different galaxy types, to select a mass-complete sample
we need to identify at any redshift the smallest observable stellar
mass for the red/early type galaxies, which have on the mean the
higher M/L at any given mass.  In practice, as proposed by
\cite{zucca2006_VVDS_LF}, we fit the Spectral Energy Distribution of
our galaxies to six templates (four observed spectra,
\citealp{CWW1980}, and two starburst SEDs, \citealp{BC1993}), and we
compute our mass completeness using those galaxies that have been
assigned the earliest template, \ie the E/S0 template. We define as
the limit mass the mass at which we are 95\% complete for these
galaxies, at any given redshift.  Figure \ref{mass_limits} shows the
mass-redshift plane of the 10k-sample, the red curve representing the
95\% mass completeness described above.

As it can be noticed, the mass limit at $z=1.0$ is very high. If we
used it in the whole range $0.1\leq z\leq1.0$ in order to select a
mass limited sample homogeneous at all redshifts, we would be left
with too few galaxies at low $z$.  We will also show that for the
analysis carried on in this part of our work we are interested in as
large a mass range as possible at any redshift. For these two reasons
it is more convenient to separately adopt the smallest limit mass we
can reach in any redshift bin. Within each $z$ bin, we keep the limit mass
constant (unlike in the case of the limiting magnitude), as it has been
shown that the Galaxy Stellar Mass Function does not evolve
significantly in the redshift range we are considering (\eg,
\citealp{pozzetti2007}).

\begin{figure}
\begin{center}
\includegraphics[width=9cm]{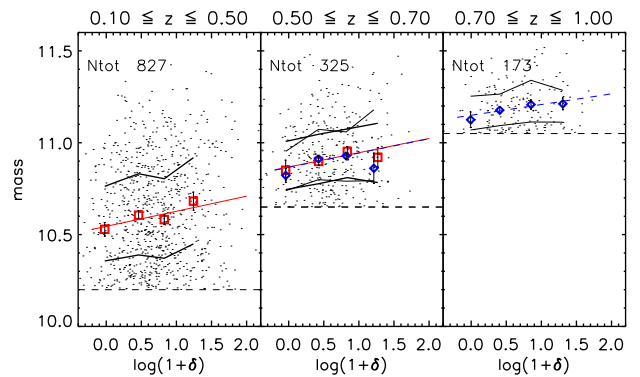}
\caption{Median stellar mass (red squares and blue diamonds) as a
function of the `count-weighted' density contrast with volume limited
tracers, for the three mass limited subsamples described in Section
\ref{mass_segr_section}. Black points are single galaxies. Red squares
are computed using the fainter volume limited tracers ($M_B \leq -19.3
-z$), blue diamonds using the bright ones ($M_B \leq -20.5 -z$).
$x$ axis values for squares and diamonds represent the
median density values in each density bin.
Vertical error bars associated to median stellar mass values are
obtained with bootstrapping technique. Black thick(/thin) lines are
the 15\% and the 85\% of the mass distribution in each density bin,
for the faint(/bright) volume-limited tracers. Horizontal dashed lines
show the mass limits in each redshift bin.  The red straight lines are
linear fits of the squares, the dashed blue lines of the diamonds, the
slopes and their errors being quoted in the text.  The total number of
galaxies considered in each redshift bin is quoted in the upper part
of each panel.}
\label{mass_segregation}
\end{center}
\end{figure}

\begin{figure}
\begin{center}
\includegraphics[width=9cm]{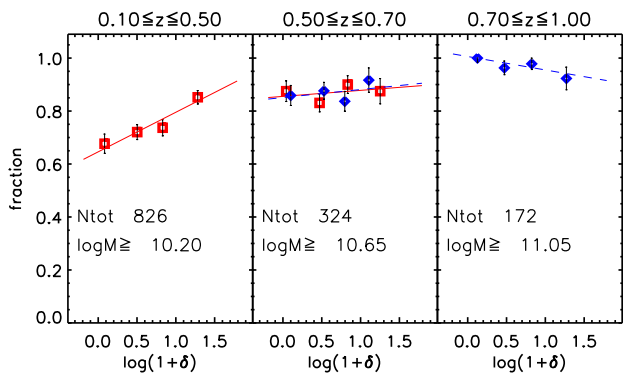}
\caption{Fraction of red ($U-B\geq1.0$) galaxies as a function
of density for three mass limited galaxy samples (mass lower limit in
the label). Red squares are computed using the fainter volume limited
tracers ($M_B \leq -19.3 -z$), blue diamonds using the bright ones
($M_B \leq -20.5 -z$). The values on the $x$ axis represent the median
density values in each density bin. The red straight lines are linear
fits of the squares, the dashed blue lines of the diamonds, the slopes
and their errors being quoted in the text.  The total number of
galaxies considered in each redshift bin is quoted in the lower part
of each panel.  }
\label{frac_col_zbins}
\end{center}
\end{figure}

To cover the largest available mass range we start our analysis
from $z=0.1$. This choice prevents us from using in the entire
redshift range the density contrast computed with the bright volume
limited tracers, as it is reliable only for $z\geq 0.4$.  In contrast,
the fainter volume limited tracers ($M_B \leq -19.3 -z$) represent a
complete sample only up to $z = 0.7$. We decided to use the fainter
volume limited tracers in the redshift interval 0.1-0.7, and the
bright one for the remaining bin $0.7 \leq z \leq 1.0$. We verified
that this choice does not introduce any spurious dependence on
redshift, because of the change, at $z=0.7$, of the tracers used to
measure the density. We took as a reference the redshift bin $0.5 \leq
z \leq 0.7$, for which we have the density computed with both the
samples, and we performed all the analyses shown in Sections
\ref{mass_segr_section} and \ref{mass_colour} with both the
tracers. We find that the results are always consistent using the two
tracers samples, as shown for example in Fig. \ref{mass_segregation}
and \ref{frac_col_zbins} (see the text below for the details about
these figures).  The same agreement is found in
\cite{bolzonella2008_MFenv}, where the galaxy stellar mass functions
in low and high density environment are computed for both tracer
types.  Therefore we believe to be safe the comparison of our analysis
at $z \leq 0.7$ and $z \geq 0.7$ using fainter and brighter tracers,
respectively.


In summary, we selected the three redshift bins [0.1-0.5], [0.5-0.7],
[0.7-1.0], with the mass limits $M_{lim}$ given by
$\log(M_{lim}/M_{\odot}) =$ 10.2, 10.65 and 11.05 respectively. 
We then studied the mass-density relation within these mass limited
samples. This is shown in Fig. \ref{mass_segregation}, where we
plot the median stellar mass (red squares and blue diamonds) as a
function of density.  Red squares are computed using the fainter
volume limited tracers, blue diamonds using the bright ones. The
density bins are chosen in the following way. For the fainter tracers,
we selected all the galaxies in the range $0.1 \leq z \leq 0.7$ with
$\log(M/M_{\odot}) \leq 10.65$ (our highest mass limit for $z\leq
0.7$), and then we divide this sample in 4 equipopulated density
bins. Then we use the so-found density thresholds in the two redshift
bins 0.1-0.5 and 0.5-0.7. Similarly, for the brighter tracers we took
all the galaxies with $\log(M/M_{\odot}) \leq 11.05$ within $0.5 \leq
z \leq 1.0$, we divide this sample in 4 equipopulated density bins,
and we use these density thresholds in the two redshift bins 0.5-0.7
and 0.7-1.0.  The red lines in
Fig. \ref{mass_segregation} are linear fits to the squares, the blue
dashed lines to the diamonds. The slope values and their errors are
$0.9\pm0.03$ and $0.8\pm0.04$ for the squares in the first and second
redshift bin, and $0.8\pm0.04$ and $0.6\pm0.03$ for the diamonds in
the second and third redshift bin, respectively.

It can be noticed that the median mass depends on environment at
$z<0.5$, but this dependence weakens for higher redshift.  This may
not be a pure time-evolution effect, because also the mass limit
changes with redshift.  Considering only the mass limited sample in
$0.1\leq z \leq 0.5$, we verified that within this sample the
mass-density relation weakens when increasing the mass limit. These
findings are in agreement with \cite{bolzonella2008_MFenv}, who show
that the GSMFs in the 10k-sample in the low and high density regions
differ mainly in the low-mass part, with this difference decreasing
for higher redshift.


\section{Are the environmental effects on galaxy colour driven by mass?}\label{mass_colour}

In the previous Sections we have shown how galaxy $U-B$ colour and
stellar mass depend on environment. It is also well known that colour
depends on stellar mass, thus we want to study whether the
environmental dependence of $U-B$ is simply the effect of the combined
colour-mass and mass-density relations. We address this issue studying
the threefold relation among density contrast, stellar mass and
colour. In the following subsections we study the colour-density
relation in the mass limited samples (Section \ref{col_den_mass_lim})
and then we discuss how to disentangle the colour-mass-environment
relation (Section \ref{col_den_mass_bins}).

\subsection{The colour-density relation in mass-limited subsamples}\label{col_den_mass_lim}

The three panels of Fig. \ref{frac_col_zbins} show the fraction of red
($U-B\geq1.0$) galaxies as a function of density for the three mass
limited samples presented in Section \ref{mass_segr_section}. Red
squares are for fainter volume limited tracers, and blue diamonds for
the brighter ones. The density bins along the $x$ axis are selected as
in Fig.  \ref{mass_segregation}. The linear fits to the squares and
diamonds are represented with solid and dashed lines, respectively.
For these fits, we find the following slopes (and $1\sigma$ confidence
level): $0.17\pm0.04$ and $0.05\pm0.05$ for the squares in the first
and second redshift bin, and $0.05\pm0.06$ and $-0.01\pm0.05$ for the
diamonds in the second and third redshift bin, respectively.  We find
that there is a clear colour-density relation within the mass-limited
sample in the lowest redshift bin, but we do not detect it at
$z>0.5$. We remind the reader that this different behaviour at
different redshift may not be a pure evolutionary effect, because in
this figure we consider samples with different lower mass limits (see
also the last paragraph of Sect. \ref{mass_segr_section}). We do not
want to make any statement about galaxy evolution inspecting
Fig. \ref{frac_col_zbins}.

\begin{figure}
\begin{center}
\includegraphics[width=9cm]{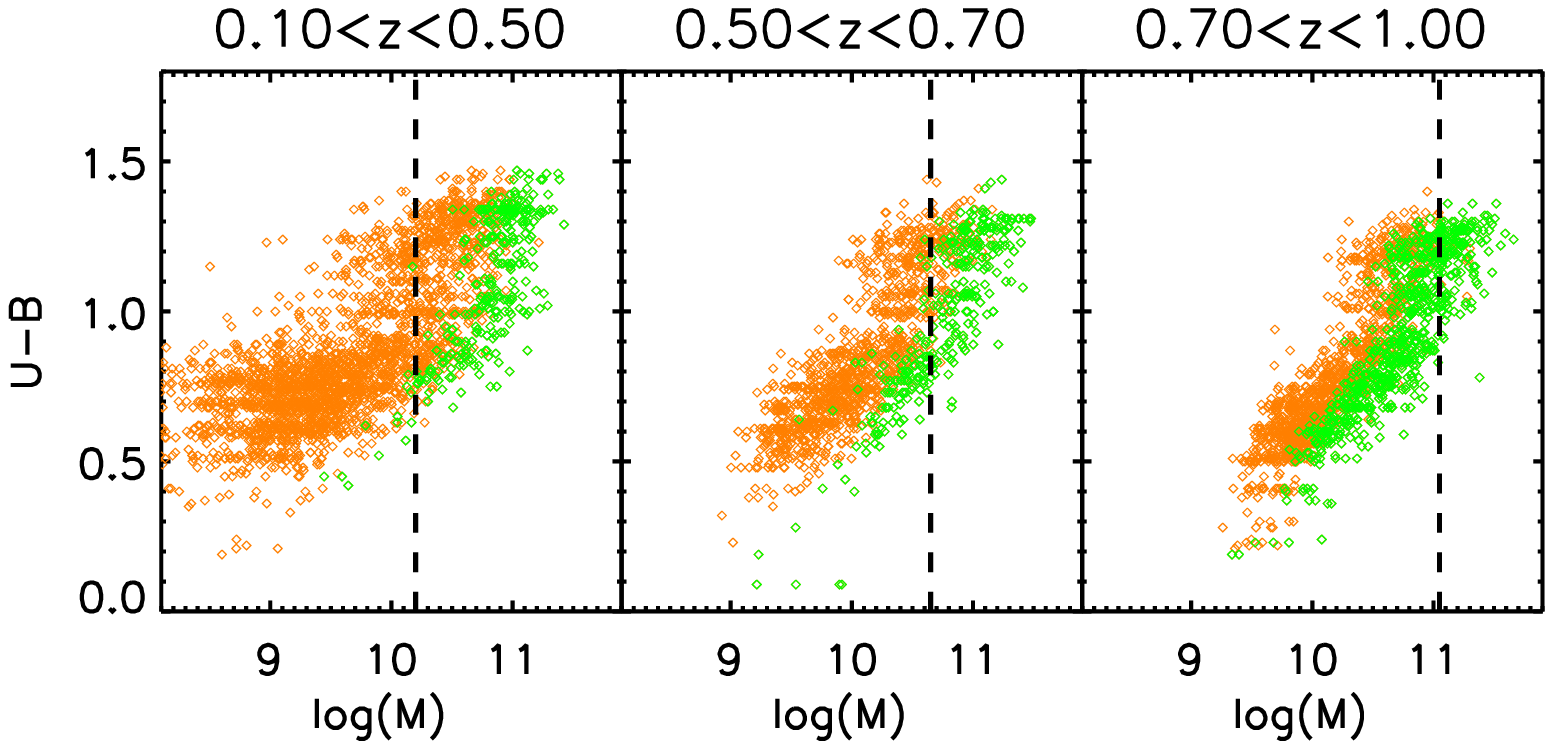}
\caption{The colour-mass plane for 3 different redshift bins. Orange
points are all the galaxies in the redshift bin considered, green
points are those galaxies that respect the luminosity limit as in Fig.
\ref{plot_frac_col_orig} and vertical lines represent the mass limit
at each redshift. }
\label{colour_mass_contours}
\end{center}
\end{figure}

It can be noticed that results presented here are quite different from
the ones in Fig. \ref{plot_frac_col_orig}, where a luminosity-selected
volume-limited sample was used and where the colour-density relation
was steeper and with a lower normalization. To understand these
differences we inspect Fig. \ref{colour_mass_contours}. For the three
$z$ bins of Fig.  \ref{frac_col_zbins}, it shows the colour-mass plane
for all the galaxies (orange points), with green points representing
the galaxies considered in the luminosity-selected volume-limited
sample of Fig. \ref{plot_frac_col_orig} ($M_B\leq -20.5-z$) and the
vertical lines the mass limits of Fig. \ref{frac_col_zbins}.

The higher normalization in Fig. \ref{frac_col_zbins} is explained by
the fact that with the chosen mass limits we select an higher fraction
of red galaxies. This effect increases with $z$, up to the highest
redshift bin, where almost only red galaxies are selected. This
observation suggests that the fixed colour cut at $U-B=1$ to define
red galaxies is not well suited for the study of our mass-selected
samples, and we will expand on this point in the next Subsection.

We also notice how the colour-density relation for the mass-selected
sample within $0.5\leq z \leq 0.7$ (middle panel in
Fig. \ref{frac_col_zbins}) is significantly flatter than the one
obtained for the luminosity-selected volume-limited sample within
roughly the same redshift interval ($0.4\leq z \leq 0.7$, first panel
of Fig. \ref{plot_frac_col_orig}). Observing the second panel in Fig.
\ref{colour_mass_contours}, we see that the explanation for this
flattening is related to the different galaxy populations that are
selected using luminosity or mass limits. In the luminosity-selected
volume-limited sample there is a population of lower-mass blue
galaxies that is not present in the mass limited sample due to their
low M/L. Moreover, this population has no red counterpart (at the same
mass) in the luminosity selected sample, as these redder galaxies are
too faint to satisfy the luminosity-selection criterion (they have on
the mean an higher M/L).

This tail of less massive blue galaxies that we find in the
luminosity-selected sample inhabits preferentially low density
regions. This can be deduced noticing that the red fractions in
Fig. \ref{plot_frac_col_orig} are on average always lower than those
in Fig. \ref{frac_col_zbins}, but the percentage difference is much
larger in the low density bins. It remains to be clarified whether
these galaxies reside mainly in low density environment only because
of their low mass (the mass segregation effect that we showed in
Fig. \ref{mass_segregation} and that is detailed in
\citealp{bolzonella2008_MFenv}), or because environment also plays a
role in shaping their colour. We will disentangle this dependence in
the following Subsection.

Here, it is worth remarking the following.  We are aware that the M/L
spread is particularly large in our sample because of the relatively
blue band-passes we use (U- and B-band). These bands are more
sensitive to dust reddening, and moreover the light at these
wavelengths is dominated by younger stars, that not only are a small
fraction of the mass, but also can be the product of transient
phenomena, such as sudden bursts of star formation. As already
explained, our choice of the B-band for the luminosity selection and
of the U-B colour for the analysis is dictated by the fact that the
I-band flux limit of our survey ($I_{AB}\leq22.5$) corresponds at the
rest frame B-band at $z\sim0.8$. This allows our sample to be not
significantly affected by colour incompleteness in the redshift range
we are interested in ($z\lesssim1$). Moreover, in a
luminosity-selected sample the M/L spread is in general unavoidable,
irrespectively of the band used for the selection, because it is
produced at least by the variety of star formation histories. This
gives rise in turn to a large spread in stellar masses, and together
with the colour-mass relation, this causes the colour-density relation
in a luminosity-selected sample to be strongly biased by the
mass-density relation. In a mass-limited sample this bias is weaker,
as the allowed mass range is smaller, but the mass-density relation
does not disappear. Thus, if we want to study the direct environmental
effects on galaxy properties, such as the colour, without being biased
by the mass segregation as a function of environment, we need to
select our sample(s) in narrow mass bins.

\begin{figure*}
\begin{center}
\includegraphics[width=15cm]{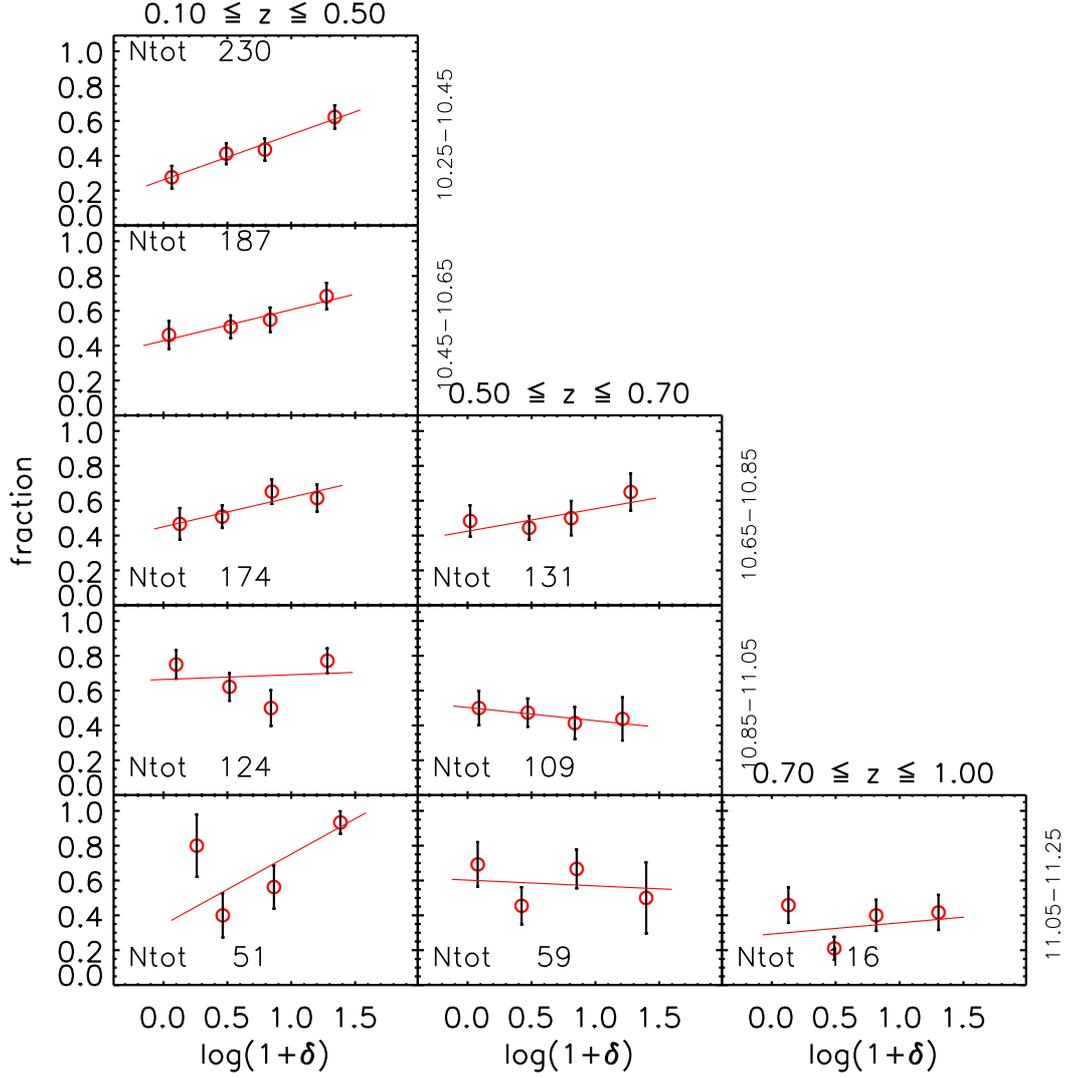}
\caption{Colour-density relation using number density and volume
limited tracers for subsamples of galaxies divided in three redshift
bins (columns, redshift bins quoted on the top) and five mass bins
(rows, mass bins quoted on the right, in logarithmic scale). For
$z\leq 0.7$ we use the fainter volume limited tracers ($M_B \leq -19.3
-z$), for $z\geq 0.7$ brighter ones ($M_B\leq -20.5 -z$). Red circles
show the fraction of galaxies with $U-B$ colour redder than a
threshold that depends on the stellar mass, as shown by the diagonal
red line in the first panel of Fig. \ref{mass_bins_running} (see also
text for details).  The values on the $x$ axis are the median
density values in each density bin. Red lines represent the linear
fit of circles.  Their slopes and associated $1\sigma$ confidence
levels are quoted in Table \ref{table_slopes_massbins}. In each panel,
`Ntot' is the total number of galaxies in that mass and redshift bin,
while the number of galaxies satisfying the colour threshold is quoted
in Table \ref{table_slopes_massbins}. }
\label{frac_col_massbins}
\end{center}
\end{figure*}

\begin{table*}
\caption{Slopes of the linear fits plotted in the panels of
Fig. \ref{frac_col_massbins}. Different rows are for the different
mass ranges, and the columns for the three redshift bins. $N_{v.r.}$
is the number of `very red' galaxies, and $N_{tot}$ the total number
of galaxies in each panel.}
\label{table_slopes_massbins}
\centering
\vspace{0.2cm}
\begin{tabular}{c c c c c c c}

\hline\hline

      Mass range & \multicolumn{2}{c}{0.1$\leq z \leq$0.5}  & \multicolumn{2}{c}{0.5$\leq z \leq$0.7}  & \multicolumn{2}{c}{0.7$\leq z \leq$1.0}\\

                 &  Slope$\pm 1\sigma $ &  $N_{v.r}/N_{tot}$ & Slope$\pm 1\sigma $ &  $N_{v.r.}/N_{tot}$ &	Slope$\pm 1\sigma $ &  $N_{v.r.}/N_{tot}$ \\

\hline
	 10.25-10.45  &  0.26$\pm$0.07  &  101/230 &  -  &  - &  -  &  -\\
	 10.45-10.65  &  0.18$\pm$0.07  &  101/187 &  -  &  - &  -  &  -  \\
	 10.65-10.85  &  0.17$\pm$0.10  &  99/174 &  0.13$\pm$0.10  &  65/131 &  -  &  -  \\
	 10.85-11.05  &  0.03$\pm$0.08  &  83/124  & -0.08$\pm$0.12  &  50/109 &  -  &  -  \\
	 11.05-11.25  &  0.41$\pm$0.12  &  33/51   & -0.03$\pm$0.16  &  33/59 &   0.06$\pm$0.11  &  42/116  \\

\hline
\hline

\end{tabular}
\end{table*}

\subsection{Disentangling the colour-mass-density relation}\label{col_den_mass_bins}

\begin{figure*}
\begin{center}
\includegraphics[width=13cm]{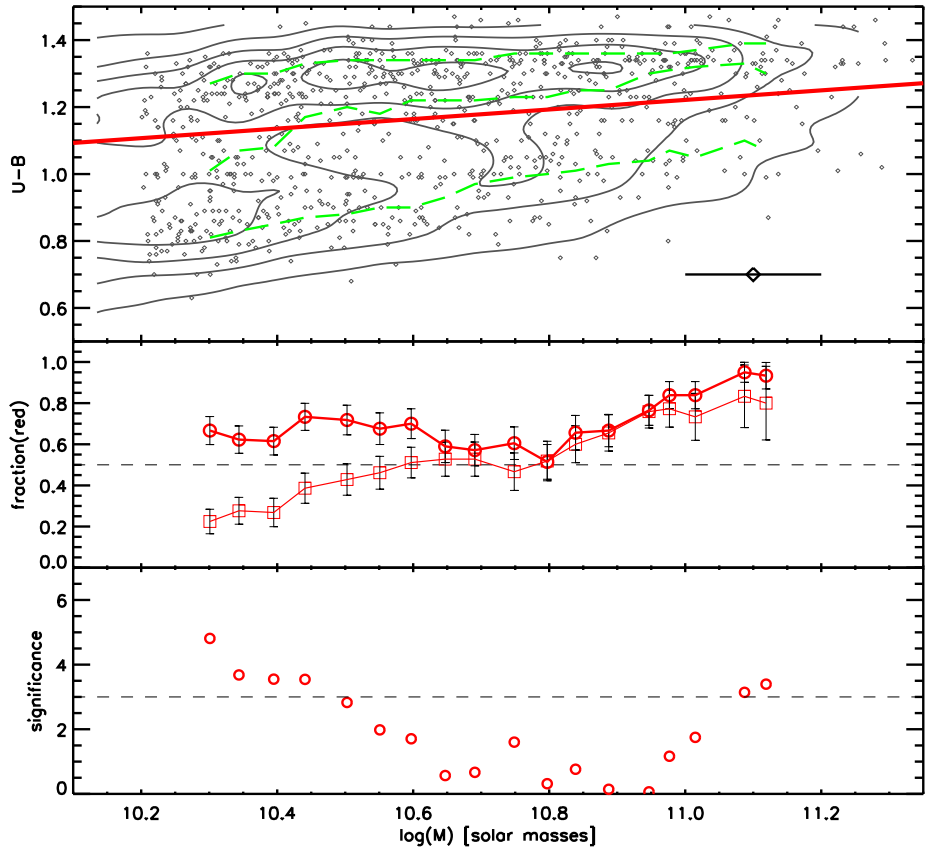}
\caption{\emph{Upper panel}: colour-mass plane for galaxies in the
redshift range $0.1<z<0.5$, above the mass limit
$\log(M/M_{\odot})=10.2$. Green dashed lines represent 15\%,50\% and
85\% of the color distribution, in bins of 0.2 in mass (logarithmic
scale, as represented by the horizontal line in the bottom right
corner). Black contours are the isodensity contours of the plotted
points. The red straight line is the colour thresholds used to define
`very red' galaxies. It is a straight cut following roughly the red
sequence and 0.3 magnitudes bluer than the red sequence itself.
\emph{Middle panel}: in not-independent bins of 0.2 in
$\log(M/M_{\odot})$ units, thick circles are the fraction of `very
red' galaxies in the highest density quartile and thin squares in the
lowest density quartile. The horizontal dashed line at fraction=0.5 is
for reference.  \emph{Bottom panel}: the significance of the slope of
the linear fit of the red fraction as a function of density, in each
mass bin. In these panels, density is computed with fainter volume
limited tracers and without mass weights.  }
\label{mass_bins_running}
\end{center}
\end{figure*}

Following the discussion detailed above, in this Section we study the
colour-density relation in mass bins, in order to understand whether
it is only the result of combining the mass-density and colour-mass
relations. First, we divide the three mass limited samples in mass
bins of $\Delta \log(M/M_{\odot})=0.2$, in order to cancel the
mass-density relation. We verified that in such narrow mass bins the
stellar mass does not depend any more on environment. Secondly, as
previously discussed, the colour cut at $U-B=1$ is not well suited to
define `red' galaxies in a sample with $\log(M/M_{\odot}) \gtrsim
10.7$, thus we introduce a new definition of `red' galaxies. We take a
colour threshold roughly parallel to the red sequence in the
colour-mass plane, and $\sim0.3$ mag bluer than the red sequence
itself. We used as reference the red sequence in the first redshift
bin of Fig. \ref{colour_mass_contours}. This colour threshold depends
on the stellar mass, as the red sequence becomes redder for higher
masses. The first panel of Fig. \ref{mass_bins_running} shows the
colour-mass plane within $0.1 \leq z \leq 0.5$. The red line is the
chosen colour threshold. As it is always redder than $U-B=1$, we
define as `very red galaxies' those galaxies with a colour redder than
this threshold.

In Fig. \ref{frac_col_massbins} we plot the fraction of `very red'
galaxies (red circles) as a function of the density contrast. The
columns are for three redshift bins as indicated on top (the same as
in Fig. \ref{frac_col_zbins}), while the rows represent different mass
bins: from top to bottom we consider the bins $10.25-10.45$,
$10.45-10.65$, $10.65-10.85$, $10.85-11.05$ and $11.05-11.25$ in
$log(M/M_{\odot})$ units, as indicated in the labels on the right of
the right-most panels.  The lowest mass considered in each redshift
bin is always equal or bigger than the mass limit at that given
$z$. For this analysis we used CWD and fainter volume limited
tracers for $z\leq 0.7$, and CWD and brighter volume limited tracers
for $z\geq 0.7$. We verified that our results in the range $0.5\leq
z\leq 0.7$ do not change using fainter or brighter volume limited
tracers. The dashed-dotted line is the linear fit of the circles
(considering their error), and `Ntot' indicates the total number of
galaxies in each panel.  Table \ref{table_slopes_massbins} shows the
fit slopes with their error, together with the number of `very red'
galaxies.

Figure \ref{frac_col_massbins} shows that the fraction of `very red'
galaxies seems generally not to depend on environment once mass is
fixed, but we can notice some exceptions. In fact, in the redshift
range $0.1\leq z \leq 0.5$ we see that the fraction of `very red'
galaxies depends on environment for $\log(M)\lesssim 10.7$. 
This fraction seems to show some trend with density also for the highest
masses explored ($\log(M/M_{\odot})\gtrsim 11.0$), but for these high
masses we have larger error bars due to the lower statistics. We note
that the same trend is observed by \cite{tasca2009}, when they study
the fraction of the more massive early type galaxies as a function of
local density.  We will explore this mass regime with higher
statistics with the complete zCOSMOS bright sample.

We addressed this issue more in detail, focusing our attention on the
redshift range $0.1\leq z \leq 0.5$, where the colour-mass-density
relation can be better studied thanks to the larger mass range spanned
above the mass limit. The aim is to analyze in a more continuous and
smooth way the colour-density relation as a function of mass bins.

The first panel of Fig. \ref{mass_bins_running} shows a zoom-in of the
first panel of Fig. \ref{colour_mass_contours}, but in this case only
above the completeness mass limit $\log(M/M_{\odot}) = 10.2$. The
black solid curves are the density contours and the green dashed lines
are the $15^{th}$, $50^{th}$ and $85^{th}$ percentiles of the colour
distribution in not-independent mass bin of $\Delta
\log(M/M_{\odot})=0.2$, each shifted from the previous one by
$\Delta_{shift}\log(M/M_{\odot})=0.05$. In this panel we show with a
red diagonal line the colour threshold used also in Fig.
\ref{frac_col_massbins} to define `very red' galaxies. 
For each mass bin, we computed the
fraction of `very red' galaxies in four density bins, defined as in
Fig. \ref{frac_col_massbins}. In the second panel of
Fig. \ref{mass_bins_running} we plotted the very red galaxy fraction
for the lowest and highest density bins, with squares and circles 
respectively. Then in each mass bin we computed a linear fit of the
four fraction values as a function of density, and in the third panel
of Fig. \ref{mass_bins_running} we plotted the significance (the
$\sigma$ level) of the slope of these fits. The horizontal dashed line
is the $3 \sigma$ significance level, for reference.

Figure \ref{mass_bins_running} more clearly shows the results of
Fig. \ref{frac_col_massbins}. If we fix the galaxy stellar mass, some
colour-density relation is still measurable at least at a $3 \sigma$
significance level for galaxies with $\log(M/M_{\odot}) \lesssim
10.6$. We note that for these low masses the galaxy population is
still composed by two main colour sequences, or clouds, divided by the
so called green valley. On the contrary, according to the colour-mass
relation, galaxies with higher masses have mainly redder colours.

The central panel of Fig. \ref{mass_bins_running} indirectly shows
also that a colour-mass relation holds irrespectively of
environment. In fact the definition of `very red' galaxies implies
that we select redder and redder galaxies for higher masses. We see
that the fraction of `very red' galaxies in high densities is almost
constant as a function of mass and this means that in this environment
galaxy colour becomes globally redder with increasing mass. The
reddening of the galaxy population for higher masses is even more
evident in low density regions, where the fraction of `very red'
galaxies increases as a function of stellar mass.

Finally, we observe that the fraction of `very red' galaxies seems to
have some (weaker) dependence on the density contrast also for the
highest masses explored. Unfortunately the number of such massive
objects in our sample is quite low (less than 100 galaxies in each of
the last three mass bins of Fig. \ref{mass_bins_running}), and so the
observed relation between density and the `very red' galaxy fraction
is not significant (if real). It will deserve a deeper investigation,
once the complete zCOSMOS \emph{bright} sample will be available
(almost doubling the statistics of the 10k-sample).


\begin{figure*}
\begin{center}
\includegraphics[width=11cm]{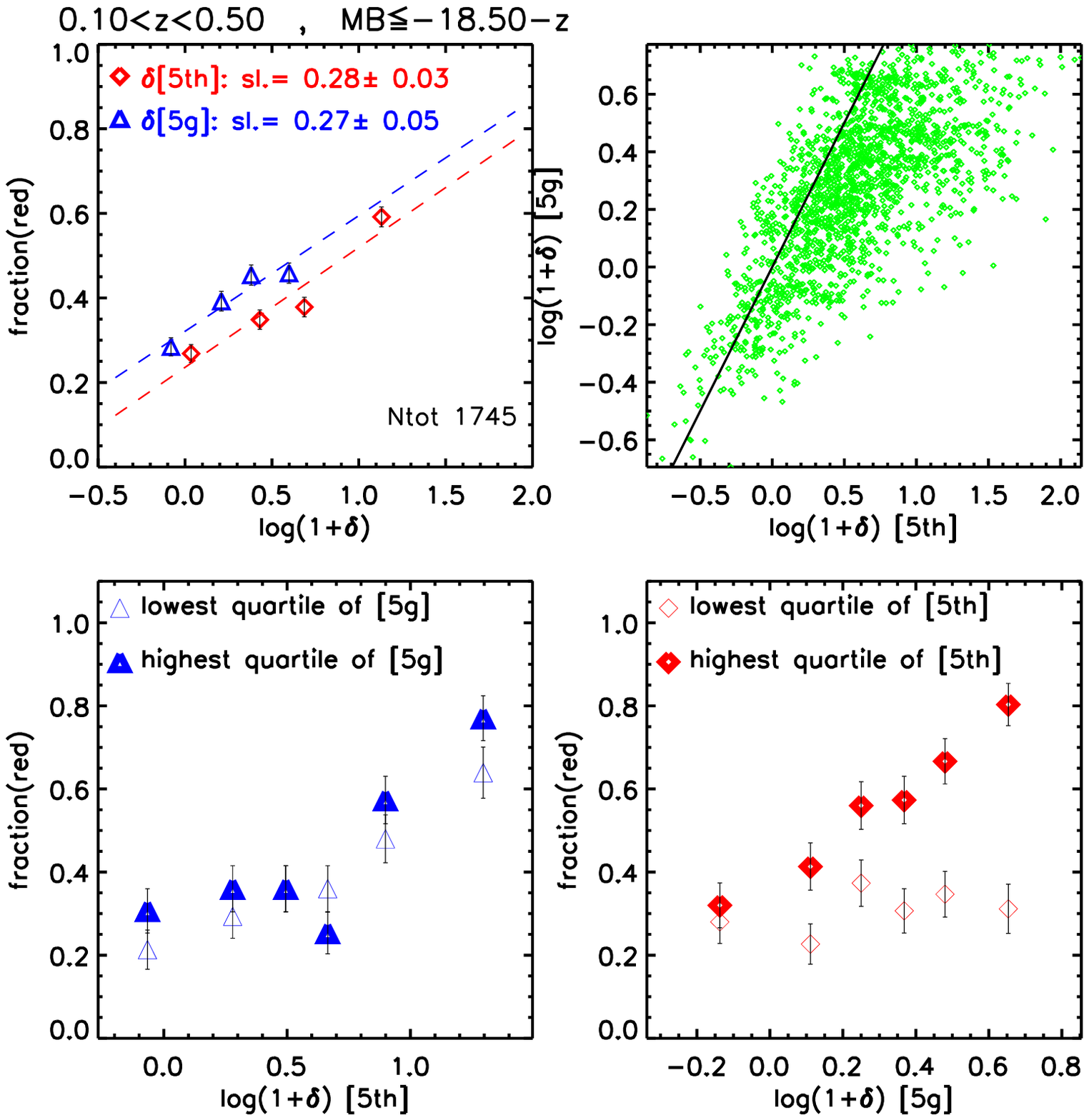}
\caption{\emph{Top-left panel}. Colour-density relation for galaxies
with $M_B \leq-18.5-z$ and $0.1<z<0.5$.  We use CWD computed with flux
limited tracers. Red diamonds are for the 5\thnn density estimator
([5th]), while the blue triangles for the density computed with a 5
Mpc Gaussian filter ([5g]).  $x$ axis values are the median value in
four roughly equipopulated bins of the density distribution, vertical
error bars are given by the usual binomial formula.  `Ntot' is the
number of galaxies in the subsample, while the linear fit slope value
is quoted with its error for both the densities.  \emph{Top-right
panel}. Scatter plot of [5th] ($x$ axis) and [5g] ($y$ axis)
densities. The straight line is the bisector. \emph{Bottom-left
panel}. For any small bin of [5th] density ($x$ axis, equipopulated
bins of 300 galaxies), we plot the fraction of red galaxies in the
first and last quartile of the [5g] density distribution (thick
symbols are for the highest density bin, light symbols for the lowest
density bin). \emph{Bottom-right panel}. The opposite of the
bottom-left panel. For any small bin of [5g] density ($x$ axis), we
plot the fraction of red galaxies in the first and last quartile of
the [5th] density distribution (same symbols as in the bottom-left
panel).}
\label{scales_dep_fig}
\end{center}
\end{figure*}


\section{Scale dependence of environmental effects}\label{scales_section}

In the previous Sections we have studied environment on the shortest
scales on which we could compute it, given the limits imposed by the
mean interparticle separation of the galaxies in our sample, and thus
the limits of density reconstruction reliability found using mock
catalogues \citep{kovac2010_density}. Moreover, we used volume
limited tracers for the reasons listed in
Sect. \ref{density_approach}, while flux limited tracers can probe
shorter scales (see Fig. 13 in \citealp{kovac2010_density}).

The issue of the scale on which density is computed is yet of great
interest. For example, a still open question is whether the effects
seen on large scales are only 'residual' of those seen at smaller
scales, or whether they add information.  Some works at $z\sim0.1$
carried out with SDSS and 2dFGRS data \citep{kauffmann2004,
blanton2006} assess that environmental dependence of galaxy properties
on scales larger than 1 $h^{-1}$Mpc are driven only by the influence
of scales $\leq 1h^{-1}$ Mpc. It could be interesting to extend this
study at higher redshift, to provide clues for any interpretation of
their results, and also possibly to add useful information for
planning future survey strategies.

We thus tried to address this issue using our available data. While
\cite{kauffmann2004} and \cite{blanton2006} used cylindrical filters
with fixed radii, within zCOSMOS 10k-sample we can not reach scales as
small as $\sim 1h^{-1}$ Mpc with a fixed aperture, as explained in
Section \ref{density}. This is expected due to the high mean
interparticle separation of the 10k-sample, that increases from $\sim
3$ $h^{-1}$Mpc to $\sim7.5$ $h^{-1}$Mpc in the range $0.1\leq z \leq
1.0$. We do reach scales smaller than $1h^{-1}$ Mpc only with the
5\thnn technique, using flux limited tracers, and only when
considering very high densities.

To be more precise, as one can see in Fig. 13 in
\cite{kovac2010_density}, using flux limited tracers and the 5\thnn
method we reach scales below $1h^{-1}$ Mpc for overdensities
($\log(1+\delta)\geq 0$) only within $z\lesssim 0.5$.

Of course using an adaptive scaling (5\thnn) as a reference is not
trivial, as scale varies with density. Nevertheless, we addressed this
issue following an approach similar to the one shown in Fig. 14 of
\cite{kauffmann2004}. This figure shows that the median value of
$D_{n}$4000 depends on the number of galaxy neighbours, on both small
($1h^{-1}$Mpc) and large (5$h^{-1}$ Mpc) scale. The question is
whether the density on the larger scale still has effects on galaxy
$D_{n}$4000 once the density on the smaller scale is
fixed. \cite{kauffmann2004} show that median $D_{n}$4000 does not
depend on large scale environment, once the small scale density is
chosen to be enclosed in a small range of very low densities or very
high densities.  We reproduce this analysis within the zCOSMOS
10k-sample.  We consider the density computed on a 5$h^{-1}$ Mpc
Gaussian filter as the large scale environment, and we use the 5 \thnn
as the density on the smaller scale, of course paying attention at the
variation of this scale itself with density. We will call these
densities [5g] and [5th] respectively. Our result is shown in
Fig. \ref{scales_dep_fig}. The Figure is divided in 4 panels.

The top left panel shows the fraction of red galaxies (U-B $\geq$ 1.0)
for the subsample of galaxies within $0.1<z<0.5$ and with
$M_B\leq-18.5-z$, as a function of the density contrast. The density
is computed with flux limited tracers, and without mass-weights, for
both the [5th] (red diamonds) and the [5g] (blue triangles). The value
of the slope of the colour-density relation is compatible within the
two density estimators (scales), but with the [5th] we reach higher
densities, where we observe the highest red fraction.

The top right panel shows the relation between [5th] and [5g]
densities, on the $x$ and $y$ axis respectively. The solid line
represents the bisector, for reference. It can be noticed that for the
lowest densities, the two environment estimators are very
similar. This is due to the fact that the distances from the 5\thnn
are getting closer to the scale of the [5g]. On the contrary, for high
densities, the [5th] density contrast reaches values well above the
maximum spanned by [5g], because of the smaller and smaller scales
probed by the [5th].

In the bottom panels we want to test whether some environment-colour
relation is still present on larger scales ([5g]), when selecting
narrow bins of densities on small scales ([5th]), and vice-versa.

The bottom-left panel of Fig. \ref{scales_dep_fig} shows the
following. We divided the [5th] density distribution in equipopulated
bins of roughly 300 galaxies ($x$ axis), and for each of these bins we
plot the fraction of red galaxies in the first and last quartile of
the [5g] density distribution (thick symbols are for the highest
density bin, light symbols for the lowest density bin). No trend with
density on large scale is left, if we fix the small scale density.

The bottom-right panel shows the opposite analysis. For any
equipopulated bin of [5g] density ($x$ axis), we plot the fraction of
red galaxies in the first and last quartile of the [5th] density
distribution in that bin. In this case, we see that a quite strong
trend with density on small scale is left, if we fix the large scale
density. This is more evident where [5th] reaches smaller scales, that
is for higher densities.

From these plots, we conclude that galaxy $U-B$ colour is affected by
small scale ($< 1$ $h^{-1}$ Mpc) environment, while it apparently
depends on the local density on larger scale only as a mirror of the
dependence on small scale. This at least it is true up to $z=0.5$, and
for a sample of galaxies with $M_B\leq -18.5 - z$.  Unfortunately it
is not possible with our data to extend this study above this
redshift, as we do not reach scales small enough ($<$1 $h^{-1}$ Mpc)
given the increase of the mean interparticle separation.


\section{Comparison with other works}\label{comp_literature}

Environmental effects on galaxy properties have already been studied
within other data sets up to $z\sim 1$ and above. Considering in
particular spectroscopic surveys, the colour-density relation has been
analyzed within the VIMOS-VLT Deep Survey (VVDS) and the DEEP2 Galaxy
Redshift Survey. Our results are qualitatively in agreement with their
findings, showing that the colour-density relation in a luminosity
selected sample weakens for increasing redshift.

More in detail, using VVDS data (a purely flux limited galaxy sample
with $I_{AB}\leq 24.0$) \cite{cucciati2006} show that galaxy colour
depends on the density contrast, with increasing strength for lower
redshifts and for brighter galaxies. They show also the absence of the
colour-density relation at $z\sim1$, and a possible reversal of the
relation itself for $z>1.2$, for luminosities around
$M_{B(H_0=70)}\leq -21.27$.  In the luminosity and redshift ranges
that we have in common, we find that our results (Fig.
\ref{plot_frac_col_orig}) are consistent with those presented in
\cite{cucciati2006}, even if we are using here a slightly different
density estimator. Our results of Fig. \ref{plot_frac_col_orig} are in
agreement also with the ones obtained with DEEP2 data, as shown by
\cite{cooper2007col} when they use a purely luminosity selected sample
(although their definition of red galaxy does depend on absolute
magnitude).

We remark that \cite{cooper2007col} obtain quite different results if
they use a second differently selected sample. They show that the
colour-density relation is still present at $z\sim1$ and disappears
only at $z\sim1.3$ if they use a luminosity-selected sample with a
colour-dependent luminosity limit, the limit being brighter
for redder galaxies. In Section \ref{col_den_mass_lim} we discussed
the risk of analyzing the colour-density relation in a luminosity
selected sample, and the colour-dependent luminosity limit used by
\cite{cooper2007col} may further decrease the control on biasing
effects. The selection that they use includes only bluer galaxies at faint
luminosities, while redder galaxies are included if they are bright. 
This means that the underlying spread in mass \vs 
luminosity becomes luminosity (or colour) dependent, producing a
non trivial effect caused by the mass-density relation.

Our results on the \emph{EW}[OII]-density relation
(Fig. \ref{plot_frac_SFR}, bottom panels) are also in agreement with
those obtained with the DEEP2 data \citep{cooper2006_envOII}, although
we find a less significant \emph{EW}[OII]-density relation within
$0.85\leq z \leq 1$.

Our findings are qualitatively in agreement also with those found by
\cite{scoville2007_lss}, based on the entire COSMOS data set, with the
density field computed using photometric redshifts.  In particular,
they find that more massive objects reside preferentially in higher
densities, where also the fraction of `Early Type' galaxies is higher,
and this is true at all redshifts explored ($0.2 \leq z \leq
1.1$). Moreover, they find that the median value of $\tau_{SF}$, defined as 
the inverse of the specific star formation rate (sSFR),  is
higher in more dense environment, and this dependence weakens for
higher redshifts. This is in agreement with our results 
on \emph{EW}[OII], that is a proxy for the sSFR
(thus $\sim 1/\tau_{SF}$).

Finally, it is interesting to compare our results about the
colour-density relation at fixed mass with similar works. In the
local universe, \cite{baldry2006_mass} find that the fraction of
galaxies in the red sequence is higher in more dense environments at
any given mass in the range $9 \leq \log(M/M_{\odot}) \leq
11$. \cite{kauffmann2004} show similar results considering the
dependence of D4000$\AA$ break and specific star formation rate on
local density at fixed stellar mass.  More in details,
\cite{baldry2006_mass} find that the difference between the fraction
of red sequence galaxies in the lowest and highest densities decreases
continuously for higher masses, as we also show in
Fig. \ref{mass_bins_running}.  We note that their definition of ``red
sequence galaxies'' implies a colour threshold that becomes redder for
brighter luminosities, \ie similar to the colour threshold we adopt,
that becomes redder for higher masses.

Considering higher redshift studies, we compared our results on the
colour-mass-density relation to the ones presented in
\cite{scodeggio2009_VVDSmass}, based on VVDS data.  While they find a
mass-density relation for mass limited samples, they do not find any
residual colour-density relation once fixing the stellar mass.
In the redshift range that we have in common whit them ($0.2\leq z
\leq 0.7$) we can explore only stellar masses higher than the ones
they study. Nevertheless, as we find a
colour-density relation at fixed mass for $10.2 \lesssim
\log(M/M_{\odot}) \lesssim 10.7$ within $0.1\leq z \leq 0.5$, while
they do not find it for slightly lower stellar masses in the same
redshift range, we studied in detail the differences between the two 
analyses.

First, we observe that their red sequence in the colour-mass plane is
much less populated then ours, with respect to the blue cloud. We
verified this inspecting directly VVDS data, while for reference we
refer the reader to Fig. 1 in \cite{franzetti2007_bimodality}, where a
colour-magnitude diagram is shown, mirroring the colour-mass plane.
De-populating randomly our red sequence in order to lower our global
fraction of red galaxies down to the value found in the VVDS sample,
we find that the fraction of the remaining red galaxies is less
dependent on environment than before.  Secondly,
\cite{scodeggio2009_VVDSmass} use a density contrast smoothed on a
larger scale with respect to the one we use. Recomputing our
colour-density relation in mass bins using the same scale they use, we
obtain a weaker colour-density relation. We conclude that the
differences between \cite{scodeggio2009_VVDSmass} and our work can be
fairly explained, and that our results are robust against these
differences.

\section{Discussion}\label{discussion}

\subsection{The evolution of the colour-density relation in a luminosity-selected sample}\label{discussion_evo}

It is nowadays well assessed that in the local universe clear
connections exist between the local environment and several galaxy
properties. On average, while redder and brighter galaxies live
preferentially in regions where the local density is higher, the
opposite is true for bluer and fainter galaxies (\eg,
\citealp{balogh2004b, kauffmann2004, blanton2005, hogg2004}).  This
picture is enriched by measurements of the two-point galaxy
correlation function for different luminosity and colour
sub-populations. More luminous galaxies tend to be more clustered that
fainter ones, and red galaxies exhibit a stronger and steeper
real-space correlation function than blue galaxies
\citep{norberg2002_clustering, zehavi2002_clustering,
madgwick2003_clustering}.  Moreover, studying the luminosity
function for early/red galaxies in different environments, a brighter
$M^{*}$ has been found for early-type galaxies residing in higher
density regions, when density is computed on several Mpc scales
\citep{croton2005}; note, however, that it is not yet clear whether
the early-type LF computed in clusters has a brighter $M^{*}$ when
compared with the LF in the field (see for example the discussion in
\citealp{boselli2006_env}).

This composite picture at low redshift is globally still valid at
higher redshift ($z\lesssim1$).  On the one hand, it has been found
that the colour-density relation persists at higher redshifts
\citep{cucciati2006, cooper2007col}. On the other hand, studies of
galaxy clustering by luminosity, by spectral type and by stellar mass
at high redshift (\eg, \citealp{coil2004b,coil2004_clustering,
pollo2006_lum,meneux2006_type,meneux2007_sm}, but see also
\citealp{meneux2009_clust_lum_zCOSMOS}) have shown that red, early
type galaxies at $z \sim 1$ are more strongly clustered than their
blue counterparts (this is also true at all redshifts up to $z \sim
1.2$). Furthermore, the clustering length increases for more massive
galaxies and for brighter ones.  This is mirrored by the LF per
environment studied in \cite{ilbert2006_LFenv}, where it is shown that
at $z \sim 1$ the brighter galaxies are preferentially populating
high-density regions, while the opposite is seen for fainter galaxies.

Although it is commonly observed that the pictures at $z\sim0.1$ and
at $z\sim1$ are linked by the progressive weakening of the
environmental effects for increasing redshift \citep{cucciati2006,
cooper2007col}, it is not still totally clear when this environmental
dependence was established.  It has even been observed the presence at
$z\gtrsim1$ of colour-density and SFR-density relations that are the
opposite with respect to the ones in the local universe
\citep{cucciati2006,elbaz2007_SFR, cooper2008sfr}. Moreover, more
light has still to be shed on the origin itself of these environmental
effects, to understand if galaxy properties are related only to the
environment in which galaxies have been born (`nature' hypothesis), or
if the environment surrounding galaxies at each epoch of their
evolution continuously plays a role in shaping their properties
(`nurture' hypothesis).

Unfortunately, the evolution of the environmental effects on galaxy
properties up to $z\sim 1$ and above has always been studied in
luminosity selected samples. We have shown in the previous Sections
that this selection may give rise to a misleading interpretation,
given the underlying mass-density relation. For example, the faster
decreasing with redshift of the red fraction in high densities than in
low densities, found in both \cite{cucciati2006} and
\cite{cooper2007col}, at a first level can be interpreted as the proof
that `nurture' is at work, differently shaping galaxy properties in
different environments. But when we take into account the mass-density
relation, these findings become easily interpretable with a biased
galaxy formation: i) more massive galaxies formed first in the highest
density peaks (\eg \citealp{marinoni2005}), and being more massive they
consumed their gas reservoir more quickly, causing the fast increase
of the red fraction with cosmic time in high densities; ii) low mass
galaxies formed later in lower density regions, and we know that they
consume their gas more slowly than their higher mass counterparts, and
this is observable in the slow (or even absent) changing with time of
the red fraction in low densities.

Our results are qualitatively in agreement with the above-mentioned
picture, when analysis is performed within luminosity-selected
samples. In Sect. \ref{red_lum_dependence}, for a
luminosity-selected volume-limited sample complete up to $z=1$, we
find that the colour-density relation weakens for increasing redshift
(Fig. \ref{plot_frac_col_orig}), this trend being supported by the
$D_{n}$4000-density and \emph{EW}[OII]-density relations (Fig.
\ref{plot_frac_SFR}), that are clearly stronger in the lowest redshift
bin explored ($0.48\leq z \leq 0.7$) than in the highest one
($0.85\leq z \leq 1.0$)

More in detailed, we also find that the fraction of red galaxies (but
also of both `passive' and `active', as defined for
Fig. \ref{plot_frac_SFR}) in the highest density bin has a stronger
variation with redshift than the corresponding fraction in the lowest
density bin.  The above mentioned discussion about the galaxy biased
formation and the mass-density relation is well suited to explain our
findings. Nevertheless, it may be the case that the faster gas
consumption in high densities is not only caused by the higher masses,
but it may be accelerated by physical processes taking place in high
density regions. It is not
possible, within our results, to distinguish which are the physical
processes really at work in the overdense regions among the ones
currently proposed (\eg ram pressure stripping of gas,
\citealp{gunn_gott1972}, galaxy-galaxy merging, \citealp{toomre1972},
strangulation, \citealp{larson1980} and harassment,
\citealp{moore1996} and so on). This prevents us from a
more detailed analysis about the possible presence of nurture effects.

Finally, we note that the redshift evolution that we find for the
colour-density relation in a luminosity-selected sample is consistent
with other parallel analyses performed within the 10k-sample.
\cite{iovino2010_groups} show that the fraction of blue galaxies
$F_{blue}$ in galaxy groups is lower than $F_{blue}$ among isolated
galaxies at any redshift in the range $0.2 \leq z \leq 1.0$, with this
difference decreasing for higher redshift.  Moreover, the fraction of
morphologically early type is always higher in higher densities, with
environment parametrized using both the density field
\citep{tasca2009} and galaxy groups \citep{kovac2010_groups}.

\subsection{The role of stellar mass in the colour-density relation }\label{discussion_mass}

Several studies indicate that, among other properties such as
luminosity and morphological parameters, colour \citep{blanton2005}
and specific star formation rate \citep{kauffmann2004} are the
properties most tightly related to environment. In the past years it
has been also shown that the galaxy stellar mass drives the star
formation history of galaxies (\eg
\citealp{gavazzi1996_massSFH,scodeggio2002_cube}), that in turn is
mainly responsible for the galaxy colour. Moreover, the stellar mass
is related on the one side to the galaxy halo mass
\citep{mandelbaum2006_halo}, and on the other side to the local
environment as determined by nearby galaxies
\citep{kauffmann2004,bundy2006,scodeggio2009_VVDSmass,bolzonella2008_MFenv}.

In particular, we refer the reader to \cite{bolzonella2008_MFenv} for
a detailed study of the environmental effects on galaxy stellar mass
within the zCOSMOS 10k-sample. They show that the Galaxy Stellar mass
Functions (GSMF) is dominated by more massive galaxies in high
densities, and that this difference between the GSMF in lower and high
density environments decreases for higher redshift. We remark that these
findings are mirrored by those related to the Luminosity Function
(LF), based on the same data \cite{zucca2009_LF}.

It follows that a non-biased analysis of environmental effects on
galaxy colour (or on other properties) can be addressed only within
complete mass-selected subsamples that span a narrow mass range.

In our work, we disentangled the three-fold dependence among galaxy
colour, mass and environment.  Globally, we do not find any
colour-density relation up to $z\sim 1$ when mass is fixed
(Fig. \ref{frac_col_massbins}), with the exception of
$\log(M/M_{\odot})\lesssim 10.7$, where we see a clear colour-density
relation (see also Fig. \ref{mass_bins_running}). As discussed in
Section \ref{comp_literature}, this is in broad agreement with similar
studies in the local universe \citep{kauffmann2004,baldry2006_mass}.

Our findings about the colour-density relation at fixed mass are
coherent with the general picture depicted by other parallel analyses
based on the zCOSMOS 10k-sample. \cite{tasca2009} and
\cite{kovac2010_groups} find the same result studying the fraction of
morphologically early type galaxies as a function of the density
contrast and galaxy group membership, respectively. The same hold for
the study of the fraction of blue galaxies in groups and in isolation
\citep{iovino2010_groups}.

From these results, it seems that in the redshift range $0.1 \leq z
\leq 0.5$ there exists a sort of threshold mass
($\log(M/M_{\odot})\sim 10.7$), below which the mix of the two broad
galaxy populations (red and blue, early and late) does still depend on
local density once stellar mass is fixed. Thus for low masses
environment affects directly galaxy properties like at least colour
and morphology. 

Accordingly to these results, we asses that the colour
depends primarily on mass, but for the low-mass regime the local
environment modulates this dependence.

First, we show that a colour-mass relation holds irrespectively of
environment. Our definition of `very red' galaxies implies a selection
of redder galaxies for increasing mass. The fact that the fraction of
`very red' galaxies in high densities is almost constant as a function
of mass (thick circles in the middle panel of
Fig. \ref{mass_bins_running}) means that in this environment galaxy
colour becomes globally redder for higher masses. This effect is even
stronger in low density regions, where the `very red' galaxy faction
increases for increasing mass.  The same is shown in
\citealp{bolzonella2008_MFenv}, studying the GSMF per galaxy types in
different environments. In parallel, this behaviour is also observed
in the Luminosity Function \citep{zucca2009_LF}.

Second, in this scenario where a colour-mass relation is embedded in
both low and high density regions, the environment seems to have a
role in shaping different colour distributions in the redshift range
$0.1 \leq z \leq 0.5$ once mass is fixed. This means that environment
does play a role independently of stellar mass. For example, in the
central panel of Fig. \ref{mass_bins_running} we see that the fraction
of red galaxies is clearly a function of the density contrast at any
given mass below $\log(M/M_{\odot})\sim 10.7$ (the possible presence
of a similar trend for $\log(M/M_{\odot})\sim 11.1$ is discussed in
Sect. \ref{col_den_mass_bins}).  Continuing the comparison with the
parallel analysis on the GSMF, also in \cite{bolzonella2008_MFenv} the
role of environment on galaxy colour per mass bin emerges clearly. The
fractional contribution of the separated blue and red galaxies GSMFs
to the global GSMF varies with environment, especially in the range of
intermediate masses ($9.5\lesssim \log(M/M_{\odot}) \lesssim 10.5$).

A possible general picture to describe these findings is the
following. Considering that the reddest and most massive galaxies have
been formed earlier (at $z>2$, having already an age of $> 1$ Gyr at
$z\sim1$, see \eg \citealp{vergani2007} and references therein), and
that moreover their Star Formation Histories (SFH) are typically
faster ($\lesssim 3$ Gyr, see \eg \citealp{gavazzi2002_SFH}) than
those of lower mass galaxies, their formation epoch and their
evolution time-scale took place on the mean before an appreciable
growth of structures (at $z\sim1$ the number of structures with
$\log(M/M_{\odot}) > 5\times10^{14}$ was $\sim100$ times lower than
today, in the concordance cosmology, see
\citealp{borgani2006_review}). We do not see evident environmental
effects on these galaxies because environment could not affect their
evolution.  On the contrary, lower mass galaxies ($\log(M/M_{\odot})
\lesssim 10.7$) not only have been formed more recently, but their SFH
are also slower. Therefore, physical processes typical of high density regions
could have modified their SFH itself. For a broad picture about the role of
mass and environment in driving galaxy evolution, obtained with SDSS
and zCOSMOS data, we refer the reader to \cite{peng2010_picture}.

This picture is in agreement with the scenario that has been already
proposed in literature where both `nature' and 'nurture' affect galaxy
properties (see \eg \citealp{kauffmann2004}, \citealp{delucia2006},
\citealp{cucciati2006}), to which we can add the dependence on mass.
First, galaxy formation is `biased' \citep{marinoni2005}, meaning that
more massive galaxies formed first in the highest density peaks, later
on followed by lower mass galaxies in less dense environment. Second,
for relatively low mass galaxies the evolution is affected by complex
physical processes that depend on the local environment. According to
this picture, the properties of a low-mass galaxy are thus affected by
the environment in which the galaxy resides at any given epoch of its
evolution, and not only at the time of its formation, as a simple
imprinting.

Unfortunately, the relatively large uncertainties in our measurements
prevent us from assessing in a precise and quantitative way the
relative roles that `nature' and `nurture' did play. We defer this
study to a future work.


\section{Summary and conclusions}\label{conclusions}

In this work we use the first $\sim10000$ spectra of the purely flux
limited zCOSMOS \emph{bright} sample ($I_{AB}\leq 22.5$) to
investigate the redshift evolution of environmental effects on galaxy
spectro-photometric properties, mainly $U-B$ colour, in the range $0.1
\leq z \leq 1.0$. We refer the reader to \cite{kovac2010_density} for
a full description of the environment parametrization via the local
density field. More in details, we want to disentangle the
multiple dependence among local density, galaxy colour and stellar
mass, in order to verify whether the environment acts directly on both
stellar mass and colour, that are related to each other. Our main
results are the following.

\begin{itemize}

\item[-] Using a luminosity-selected volume-limited sample ($M_B\leq
-20.5-z$) we find, confirming previous results, that the fraction of
red ($U-B \geq 1$) galaxies depends on environment at least up to
$z\sim1$, with red galaxies residing preferentially in high density
environments. This trend becomes weaker for higher redshifts, and it
is mirrored by the one we observe for the fraction of galaxies with
$D_{n} 4000\geq 1.4$. We also find that the fraction of galaxies with
$\log(EW[OII])\geq 1.15$ is higher for lower densities, and this is
true up to $z\sim1$. Also in this case we observe a weakening of the
environmental effects for increasing redshift.

\item[-] Given the fact that stellar mass depends on local density
(see \eg \citealp{bolzonella2008_MFenv} and our
Fig. \ref{mass_segregation}), and given that colour depends on mass,
it is difficult to interpret the meaning of the colour-density
relation in a luminosity-selected sample.  In fact, the wide spread in
mass-to-light ratios embedded in this selection leave us with a broad
range in stellar masses, that biases any possible direct
colour-density relation via the mass segregation as a function of
environment.

\item[-] We disentangle the colour-mass-density relation using galaxy
subsamples enclosed in narrow mass bins, to avoid any dependence of
mass on local density. We study the colour-density relation in mass
bins of $\Delta \log(M/M_{\odot})=0.2$, in three different redshift
bins within the range 0.1-1.0, paying attention to attain to the mass
limits imposed by the flux limit of our survey. We find that , once
the mass is fixed, the colour-density relation is globally flat, but
with some exceptions. We observe that within $0.1 \leq z \leq 0.5$ the
fraction of `very red' galaxies depends on environment even when mass
is fixed, at least for $\log(M/M_{\odot}) \lesssim 10.7$ (see Section
\ref{col_den_mass_bins} for the definition of `very red'
galaxies). This means that environment affects directly not only the
stellar mass, but also other galaxy properties, at least for these
given mass and redshift ranges.

\end{itemize}

According to these results, we suggest a scenario in which the
colour depends primarily on mass, but for a relatively low-mass regime
($10.2 \lesssim \log(M/M_{\odot}) \lesssim 10.7$) the local
environment modulates this dependence. These galaxies have been formed
more recently, in an epoch where evolved structures were already in
place. Moreover, the physical processes typical of high density
regions could operate during longer periods of time in shaping the
properties of these galaxies, thanks to their longer (on average) 
star formation histories.

It is not possible with this analysis to quantitatively assess the
relative role of biased initial conditions (more massive galaxies form
first and in more dense environments) and of physical processes acting
during galaxies lifetimes, in order to precisely explain the origin
(or the absence) of both colour- and mass-density relations. We
refer the reader to \cite{peng2010_picture} for a more comprehensive
analysis on how mass and environment affect galaxy evolution.  On the
one hand, the forthcoming zCOSMOS $20k$ bright survey will allow us a
more detailed analysis of these mass and redshift regimes, on the
other hand this work has to be complemented by the study of the
redshift evolution of the less massive galaxies population, that is
not possible within the zCOSMOS bright sample, and that is of crucial
importance to better understand how environment affects the different
galaxy populations.


\begin{acknowledgements}

We thank the referee for helpful comments. This work has been supported in 
part by the grant ASI/COFIS/WP3110 I/026/07/0.

\end{acknowledgements}




\bibliographystyle{aa}
\bibliography{biblio}

\end{document}